\documentclass[12pt,preprint]{aastex}   

\newcommand{\Lsol}{\mbox{$L_\odot$}}
\newcommand{\Msol}{\mbox{$M_\odot$}}

\newcommand{\asec}{\mbox{$''$}}
\newcommand{\kms}{\mbox{km s$^{-1}$}}
\newcommand{\ms}{\mbox{m s$^{-1}$}}
\newcommand{\perbeam}{\mbox{beam$^{-1}$}}
\newcommand{\persquarecm}{\mbox{cm$^{-2}$}}

\newcommand{\peryr}{\mbox{yr$^{-1}$}}
\newcommand{\persquarearcsec}{\mbox{arcsec$^{-2}$}}
\newcommand{\persquarepc}{\mbox{pc$^{-2}$}}
\newcommand{\percubicpc}{\mbox{pc$^{-3}$}}

\newcommand{\Halpha}{\mbox{H$\alpha$}}
\newcommand{\minus}{\mbox{$-$}}

\newcommand{\twelveCO}{\mbox{$^{12}$CO}}
\newcommand{\thirteenCO}{\mbox{$^{13}$CO}}
\newcommand{\CeighteenO}{\mbox{C$^{18}$O}}

\newcommand{\thirteenC}{\mbox{$^{13}$C}}
\renewcommand{\tilde}{\mbox{$\sim$}}

\newcommand{\Htwo}{\mbox{\ion{H}{2}}}
\newcommand{\HH}{\mbox{H$_2$}}

\shorttitle{IMAGING MOLECULAR GAS IN NGC 3256}
\shortauthors{Sakamoto, Ho, \& Peck}

\slugcomment{Accepted for publication in ApJ}

\begin{document}
\title{Imaging Molecular Gas in the Luminous Merger NGC 3256 : 
Detection of High-Velocity Gas and Twin Gas Peaks in the Double Nucleus
}
\author{Kazushi Sakamoto\altaffilmark{1,2}, 
Paul T. P. Ho\altaffilmark{3,4}, 
and
Alison B. Peck\altaffilmark{1}
}
\altaffiltext{1}{Harvard-Smithsonian Center for Astrophysics,
Submillimeter Array, 645, N. A'ohoku Place, Hilo, HI 96720}
\altaffiltext{2}{National Astronomical Observatory of Japan,
Mitaka, Tokyo 181-8588, Japan.  email: sakamoto.kazushi@nao.ac.jp}
\altaffiltext{3}{Harvard-Smithsonian Center for Astrophysics,
60 Garden Street, Cambridge, MA 02138}
\altaffiltext{4}{Academia Sinica, Institute of Astronomy and Astrophysics, 
P.O. Box 23-141, Taipei 106, Taiwan}

\begin{abstract}
Molecular gas in the merging starburst galaxy NGC 3256 has been imaged
with the Submillimeter Array 
at a resolution of $1'' \times 2''$ (170 $\times$ 340 pc at 35 Mpc).
This is the first interferometric imaging of molecular gas in the most luminous 
galaxy within $z$=0.01.
There is a large disk of molecular gas ($r > 3$ kpc) in the center of the merger
with a strong gas concentration toward the double nucleus. 
The gas disk  having a mass of  \tilde$3\times 10^{9}$ \Msol\ in the central 3 kpc
rotates around a point between the two nuclei that are 850 pc apart on the sky.
The molecular gas is warm and turbulent and shows spatial variation of the intensity ratio between
CO isotopomers.
High-velocity molecular gas is discovered at the galactic center.
Its velocity in our line of sight is up to 420 \kms\ offset from the systemic velocity 
of the galaxy; the terminal velocity is twice as large as that due to the rotation of the main gas disk.
The high-velocity gas is most likely due to a molecular outflow from the gas disk,
entrained by the starburst-driven superwind in the galaxy. 
The molecular outflow is estimated to have a rate of \tilde 10 \Msol\ \peryr\ and 
to play a significant role in the dispersal or depletion of molecular gas from the galactic center.
A compact gas concentration and steep velocity gradient are also found around each of the twin nuclei.
They are suggestive of a small gas disk rotating around each nucleus.  
If these are indeed mini-disks, their dynamical masses are \tilde$10^{9}$ \Msol\ within a radius of 170 pc.
\end{abstract}

\keywords{ 
        galaxies: starburst ---
        galaxies: ISM ---
        ISM: jets and outflows ---
        galaxies: individual (NGC 3256) ---
        galaxies: interactions ---
        galaxies: evolution         
       }

\section{Introduction}
The luminous infrared galaxy NGC 3256 has an infrared luminosity of  $10^{11.56}$ \Lsol\ 
at the distance of 35.4 Mpc \citep[1\arcsec=170 pc;][]{Sanders03}, which
makes it the most luminous galaxy within $z$=0.01 \citep{Sargent89}.
The vast luminosity,  largely emitted in the far-IR, comes from 
within the central $\lesssim$ 20\arcsec\ of the galaxy \citep{Smith96}.
The center of the merging galaxy hosts a starburst seen across a wide range of wavelengths
\citep[e.g.,][]{Graham84, Doyon94}.
High-resolution imaging with the Hubble Space Telescope (HST) revealed hundreds of bright young clusters
in the galactic center \citep{Zepf99, Alonso-Herrero02}.

The extreme starburst is expected to blow interstellar gas out of the system 
and may leave a gas-depleted elliptical galaxy \citep{Graham84}.
Indeed, an outflow of interstellar medium, or superwind, has been detected.
Its evidence includes 
LINER-like optical line ratios as well as large line widths (up to  6000 \kms) off the nucleus \citep{Moran99},
blueshifted absorption and emission lines \citep{Heckman00, Lipari00, Lipari04}, 
and the pattern of optical polarization (i.e., reflection by entrained dust) around 
the galaxy \citep{Scarrott96}.
These observations suggest that the galaxy wind extends out to several kiloparsecs.

NGC 3256 is in the late stage of galaxy merging \citep{VV59, Toomre77}.
There is a pair of long tidal tails seen in the optical (see Fig. \ref{fig.opt}) and in \ion{H}{1} emission,
extending 40 kpc on each side \citep{deVaucouleurs61, English03}.
Such morphology suggests a prograde-prograde merger of two gas-rich spiral galaxies of
similar size \citep{Toomre72, White79}. 
It also suggests that we view the system with a low inclination angle \citep{Feast78, English03}.
The radial light profile of the galaxy in the $K$-band is close to but has not achieved the
de Vaucouleurs profile (i.e.,  $I(r) \propto \exp(-k r^{1/4})$ ) 
of elliptical galaxies \citep{Moorwood94, Rothberg04}, unlike the majority of merger
remnants surveyed by \citet{Rothberg04}. 
This suggests that the merging of the stellar systems in NGC 3256 is not yet complete.

Likely corresponding to the incomplete merger,
two nuclei with \tilde5\arcsec\ (850 pc on the sky) separation 
have been detected in the near infrared, radio, and X-rays in the starbursting center of the merger
\citep{Moorwood94, Norris95, Lira02}.
They are aligned in the north-south direction, and 
the southern nucleus is highly obscured, rendering it invisible in the optical.
\citet{Neff03} suggested a low luminosity AGN in each nucleus from comparison of radio and X-ray observations,
while \citet{Jenkins04} found no strong evidence for an AGN in X-ray data alone.

There is abundant molecular gas (\tilde$10^{10}$ \Msol) in the galactic center around the double nucleus,  
probably feeding the starburst \citep{Sargent89, Casoli91, Aalto91a, Garay93}. 
NGC 3256 is one of the first galaxies in which unusually large \twelveCO/\thirteenCO\ intensity ratio
characteristics of luminous mergers was observed; it was attributed to the merger and starburst
environment \citep{Aalto91b,Casoli92b}.

The observations so far are broadly in agreement 
with the previous observations of other luminous and ultraluminous mergers 
and with the galaxy evolution models involving merging.
They suggest that a galaxy collision and subsequent merger 
bring most of gas in the progenitor galaxies to the merger center, 
cause a burst(s) of star formation as well as galaxy wind driven by the starburst, 
and eventually make an elliptical galaxy, sometimes with a phase of luminous nuclear activity 
\citep[and references threin]{Schweizer86, Sanders96, Genzel00, Sanders04, Veilleux05}.

The proximity of a source of this luminosity and the near face-on configuration 
make NGC 3256 an ideal target to test and refine the scenario of merger-induced starburst and galaxy evolution.
Moreover, NGC 3256 is arguably the southern-sky counterpart of Arp 220, the archetype of 
the infrared-luminous merging galaxies \citep{Soifer84}.
The two galaxies not only share the title of the most luminous galaxy within their respective distances
but also the  close binary nuclei, 
since Arp 220, which is twice as distant as and four times more luminous than NGC 3256,
has two nuclei only 0.3 kpc apart on the sky.
Thus they are most likely near the final stage of galaxy merger 
when the merger-induced evolution is expected to be most rapid and prominent.
Only the southern location of NGC 3256 (Dec. = $-44\arcdeg$) has delayed
one of the most important observations required to better understand the merger-starburst evolution
--- high-resolution observations of its molecular gas.

We have made high-resolution observations of the cold interstellar medium 
in NGC 3256 in 1.3 mm using the Submillimeter Array (SMA)\footnote{
The Submillimeter Array is a joint
project between the Smithsonian Astrophysical Observatory and the
Academia Sinica Institute of Astronomy and Astrophysics, and is
funded by the Smithsonian Institution and the Academia Sinica.
}.
Our observations aim to uncover the distribution, kinematics, and physical
properties of the cold molecular gas and dust in the merger-starburst
environment of NGC 3256 with a high resolution and sensitivity.
Of particular interest are the distribution and dynamics of gas around the 
starbursting double nucleus
and  how the molecular gas is affected by the starburst. 
The new data of \tilde1\arcsec\ resolution are from the first interferometric observations
of molecular gas in this galaxy. They provide more than 10-fold improvement in spatial resolution 
over previous single-dish observations made with $\gtrsim 20\asec$ beams.

We begin by introducing our SMA observations (\S\ref{s.observations}), 
then portray the overall distribution and kinematics of molecular gas in the central \tilde9 kpc (\S\ref{s.overall}),
report the detection of gas concentrations associated with the two nuclei and 
gas kinematics suggestive of rotation around each (\S\ref{s.double}),
report the discovery of high velocity gas and model it as a molecular outflow from the starburst (\S\ref{s.high-velocity-gas}),
and characterize the properties of molecular gas in the galactic center on
the basis of intensity ratios of CO isotopomers (\S\ref{s.gas-properties}).
Our observations are discussed in the context of the galaxy merger and merger-induced starburst
in \S\ref{s.discussion}, and summarized in \S\ref{s.conclusions}.

In this paper, we refer to the north and south nucleus as N and S, respectively,
and adopt the following positions for them on the basis of radio and X-ray observations
\citep{Neff03,Norris95,Lira02};
$\alpha (N) =10^{\rm h}27^{\rm m}51\fs23$, 
$\delta (N)  =-43\arcdeg54\arcmin14\farcs0$ 
and 
$\alpha (S) =10^{\rm h}27^{\rm m}51\fs22$, 
$\delta (S) =-43\arcdeg54\arcmin19\farcs2$ 
in J2000.
The systemic velocity of the galaxy in the literature 
is $V_{\rm sys}$ (LSR, radio) $\approx 2775$ \kms\ 
\citep{Casoli91, Aalto91a, Garay93, English03}.

\section{SMA Observations and Data reduction \label{s.observations}}
We observed NGC 3256 at 1.3 mm using the Submillimeter Array (SMA),
which consists of eight 6 m-diameter antennas at the summit of Mauna Kea, Hawaii
\citep{Ho04}.
We chose the center position of our observations to be
$\alpha=10^{\rm h}27^{\rm m}51\fs22$, 
$\delta=-43\arcdeg54\arcmin19\farcs2$ (J2000),
which is centered between the two nuclei.
The primary beam of the SMA antennas has the FWHM size of 52\arcsec\ (= 9 kpc) at 230 GHz.
The extended and compact array configurations were used for our
observations in February 20th and April 4th 2004, respectively, with
all eight antennas participating on both nights.
They provided projected baselines ranging from 6 m to 179 m.
The total integration time on the galaxy was 6.9 hours in two tracks.

The receivers  were tuned to simultaneously observe the three CO
lines, \twelveCO(2--1) in the upper side band (USB) and 
\thirteenCO(2--1) and \CeighteenO(2--1) in the lower sideband (LSB).
The center frequency of the USB was 227.720 GHz, 
and the LSB was 10 GHz lower.
Each sideband had 2 GHz of bandwidth and a spectral resolution of 
0.81 MHz.
The weather was excellent on both nights.  
The 225 GHz zenith opacity measured at the neighboring Caltech Submillimeter Observatory 
was 0.05 and 0.08 for the first and second nights, respectively. 
The double-sideband system temperature toward the galaxy was 160 K in median,
even though the galaxy was observed between the elevation of 14\degr\ and 26\degr.

We observed  the quasar J1037\minus295 once every 20 min  for gain calibration.
The flux density of the quasar was estimated to be 0.86 Jy and 0.84 Jy for LSB and USB, respectively, 
from comparison with Mars, for which we assumed the
brightness temperature of 200 K. 
The system passband was calibrated by observing Jupiter, Saturn, and a few bright quasars.
The elevation-dependent attenuation by the atmosphere was corrected
by using the system temperature.
The precision of our gain calibration was assessed by comparing, between the two tracks,
the total \twelveCO\ flux  detected in the same range of hour angle on the common baselines 
in the two array configurations.
The flux agreed within 10\%.
We therefore assign, conservatively, $\pm 10$\% for the random error of our flux measurements
reported in this paper.
The error in our absolute flux scaling, however, can be larger due to 
unknown or uncharacterized systemic effects.

The SMA data were reduced using MIR, which is an IDL version of MMA \citep{Scoville93}, 
MIRIAD \citep{Sault95}, and the NRAO AIPS package \citep{Bridle94}. 
After the standard passband and gain calibration, the 
spectral channels that did not have emission lines were combined
to make continuum data of 0.33 GHz bandwidth in each sideband.
Channels containing the high-velocity emission that we discovered were excluded from the continuum data.
The continuum was subtracted from spectral line data in the $uv$ plane.
Images of the line and continuum emission were then made by Fourier transforming the 
respective set of visibilities.
Various spatial and spectral resolutions were achieved by changing $uv$-weighting and channel binning.
Among our datasets are the lowest-resolution one made with natural weighting 
and 20 \kms\ resolution to detect faint extended emission, 
and the highest resolution \twelveCO\ maps made with the super-uniform weighting
and 10 \kms\ resolution.
Intermediate resolution maps with the robust $uv$ weighting are also used.
The rms noise in the natural weighting data cubes is $20 \pm 2$ mJy \perbeam. 
The SMA maps presented in this paper are not corrected for the primary beam attenuation,
except for Fig \ref{fig.cor-2s} (d).
The attenuation is approximately Gaussian and is a factor of 2 for a source 26\arcsec\ from the map center.
The correction for it is made in all flux and spectrum measurements, though
it is rather small for most emission that we detect.
Velocities in this paper are measured with respect to the local standard of rest (LSR) and are defined
in the ratio convention. 

We compared our data cubes with single-dish observations in the literature, and found
that our interferometric observations recovered almost all of the flux in the region.
The observations at the 15 m Swedish-ESO-Submillimeter-Telescope (SEST) 
gave the total \twelveCO(2--1)
flux of (3.2 -- 3.6)$\times 10^{3}$ Jy \kms\ and \thirteenCO(2--1) flux of (1.3 -- 1.5)$\times 10^2$ Jy \kms\
in their beam at the galactic center \citep{Casoli91, Aalto91a, Garay93}\footnote{
Only the \twelveCO(2--1) flux of \citet{Aalto91a} is cited here
because their \thirteenCO(2--1) flux is more than twice larger than the other two observations.}.
We measured total \twelveCO(2--1) and \thirteenCO(2--1) flux of 4.1$\times 10^3$ Jy \kms\
and 1.4$\times 10^2$ Jy \kms\ from our natural-weighting data corrected for the primary beam attenuation
and convolved to the SEST resolution, which we assumed to be 23\arcsec\ in FWHM at 230 GHz.
The high fraction of flux recovery is consistent with the compact size of the CO emission,
whose half-power diameter is estimated to be about 10\arcsec\  by \citet{Aalto91a}.
The 20\% larger flux in \twelveCO\ at the SMA than at the SEST may be 
due to a larger-than-usual error in the SMA calibration for this source at very low elevations, 
and may be also due to pointing and calibration uncertainties in the single-dish observations\footnote{
The size of the SEST beam given in the papers ranges from 22\arcsec\ at 1.3 mm 
(=230.6 GHz) to 25\arcsec\ at 230 GHz. 
The conversion factor from temperature to flux density scales with the square of the beam size
for a compact source.
Thus the single-dish flux values given above increase by 18\% if we adopt the largest beam size.}.
The single-dish observations must also have slightly underestimated the CO flux
by over-subtracting the emission as a linear baseline.
This is because the CO line is wider than previously believed as we see below.
However, this leads to only  1--2\% underestimation of the single-dish CO flux.
In any case, our relative calibration between the CO lines and continuum should be
more accurate than their absolute calibration, because they were simultaneously observed
with the same receivers and went through the same signal path and reduction procedure.
Regarding missing flux in our higher resolution datasets, 
our highest-resolution \twelveCO\ data cube contains 54\% of the total  flux 
in our natural-weighting data.

\section{Overall Gas Distribution and Kinematics  \label{s.overall}}
\subsection{spatial distribution of molecular gas and dust  \label{s.overall-spatial} }
The \twelveCO(2--1) emission as shown in Fig. \ref{fig.naturalmap} is highly concentrated 
toward the galactic center.
It has a half-maximum radius of about 5\arcsec\ (\tilde1 kpc) 
in the galactic plane (Fig. \ref{fig.12co.iring}).
The compact CO morphology is consistent with the SEST mapping result of \citet{Aalto91a}.
The CO central peak is elongated in the north-south direction and connects the double nucleus
of 5\arcsec\ separation.
There are two peaks in this structure in the \thirteenCO(2--1) map also shown in Fig. \ref{fig.naturalmap}.
Each \thirteenCO\ peak is spatially associated with one of the two nuclei.
We discuss these double peaks in more detail in the following sections using higher resolution images.
The gas distribution extends at least to the radius of 3 kpc in our data, 
and it may be extended beyond our field of view.
In the outer area, there are arcs or hints of spiral arms in the molecular gas.
Some of them correspond to spiral arms or dust lanes in the optical images in the literature. 
For example, the horizontal feature near the northeast edge of the \twelveCO\ map, around the offset
of $\Delta$R.A.=10\arcsec\ and $\Delta$Dec.=+15\arcsec, corresponds to a spiral arm emanating from the
galactic center \citep[see the \Halpha\ image of ][Fig. 2(b)]{Moorwood94}.

The 1.3 mm continuum and \CeighteenO(2--1) emission are also detected in the galactic center
(see Fig. \ref{fig.naturalmap}). 
The spatial distribution of the continuum is quite similar to that of \thirteenCO(2--1) emission.
There are peaks at or near the two nuclei, surrounded by extended emission.
A weaker peak is also seen \tilde6\arcsec\ east of the nucleus N, as is seen in \thirteenCO(2--1)
and \twelveCO(2--1).
The shape of the \CeighteenO(2--1) emission looks somewhat different 
from those of other emission mentioned above.
However, this is likely due at least partly to the low signal-to-noise ratio in the data.

The total \twelveCO(2--1) fluxes within the galactocentric radii ($r_{g}$) of 10\arcsec\ and 20\arcsec\
are $3.4\times 10^3$ Jy \kms\ and $5.2\times 10^3$ Jy \kms, respectively, 
as shown in Fig. \ref{fig.12co.iring}.
The total fluxes of \thirteenCO(2--1) and \CeighteenO(2--1) emission
within the galactocentric radius of 10\arcsec\ are $1.5 \times 10^2$ Jy \kms\  and  31 Jy \kms,
respectively. 
The total flux density of the 1.3 mm (222.7 GHz) continuum in the same area is 79 mJy.
Each line flux is integrated over the same 600 \kms\ range from 2530 \kms\ to 3130 \kms,
and each flux measurement for both line and continuum 
was made in the same way as explained in the caption of Fig. \ref{fig.12co.iring}.
Each flux has a 10\% scaling uncertainty as noted in \S \ref{s.observations}.

\subsection{gas mass estimate \label{s.gas-mass-estimate} }
The mass of molecular gas ($M_{\rm mol}$) in the region is estimated in various ways from the line emission.
The molecular gas mass of  $2\times 10^{10}$ \Msol\ for $r_{g} \leq 10\asec = 1.7$ kpc is
obtained from the conversion factor 
$N_{\rm H_2}/I_{\rm CO(1-0)}= 3.0\times 10^{20}$ \persquarecm\ (K \kms)$^{-1}$,
which is based on the virial analysis of molecular clouds in the disk of
our Galaxy \citep{Scoville87, Solomon87}.
We assume that the CO(1--0) and CO(2--1) lines have the same
brightness temperature and emitting area. 
This mass estimate and the ones in the rest of this section include the
contribution of helium.
The $M_{\rm mol}$ in the same region is estimated to be lower, 
$9\times 10^{9}$ \Msol\ and $2\times 10^{9}$ \Msol, from the
conversion factors derived from $\gamma$-ray observations to count H-nuclei;
the two estimates are from the conversion factor of 
$1.56\times 10^{20}$ \persquarecm\ (K \kms)$^{-1}$ assumed to be constant over the Galaxy \citep{Hunter97}
and $0.4\times 10^{20}$ \persquarecm\ (K \kms)$^{-1}$ at $r_{g}= 2$ kpc  in our Galaxy \citep{Strong04}, respectively.
Another estimate is from the \CeighteenO(2--1) emission and the prescription of
\citet{Stutzki90} and \citet{Wild92}. 
They suggested that this optically thin line should have nearly a constant
emissivity per molecule in the range of temperature and density relevant for starburst galaxies such as M82.
NGC 3256, also being a starburst, is probably in the class of galactic nuclei to which the
method is applicable.
The observed \CeighteenO(2--1) flux gives the molecular gas mass of the $1.4\times 10^{9}$ \Msol\
for $r_{g} \leq 1.7$ kpc
for the assumed \CeighteenO\ abundance of $[\CeighteenO] /[\HH]= 10^{-6.8}$ \citep{Frerking82}.
The mass of atomic gas within this radius is estimated to be $M_{\rm atom}$=(0.4--1.9)$\times 10^9$ \Msol\ by scaling
the mass estimate by \citet{English03}, who obtained the mass of HI absorbing gas in the central 23\arcsec.

The 1.3 mm continuum is also used for the mass estimate in the following way. 
First, about a half of the 1.3 mm flux density, or 45 mJy, is attributed to dust emission by subtracting
the contributions from synchrotron and free-free emission from the observed flux density.
The synchrotron emission at 1.3 mm is estimated
to be 15 mJy by extrapolating a power law (of a spectral index of $-0.76$) obtained
by fitting the radio flux measurements between 160 MHz and 5 GHz  
in the NASA Extragalactic Database\footnote{
The NASA/IPAC Extragalactic Database (NED) is operated by the Jet Propulsion Laboratory, 
California Institute of Technology, under contract with the National Aeronautics and Space Administration.}.
The free-free flux density is estimated to be 19 mJy from the \ion{H}{2} region model of \citet{Roy05}, 
who derived properties of the \ion{H}{2} regions in the galactic center by observing radio recombination lines
in NGC 3256.
After subtracting these contributions, the remaining emission of 45 mJy  must be the thermal emission 
from dust in the interstellar medium. 
We used the dust emissivity of \citet{Draine84} and the dust temperature of 44 K \citep{Smith96} to obtain the
gas (molecular plus atomic) mass of $2.6 \times 10^9$ \Msol\ for $r_{g} \leq 1.7$ kpc.
The mass is reduced by half if we adopt the dust opacity coefficient of \citet{Hildebrand83}.
An assumption made in these estimates is that the gas to dust ratio and dust properties are the same in
our Galaxy and NGC 3256.

The large scatter among the mass estimates is not surprising for a starburst nucleus.
Many assumptions that can not be easily verified are inevitably made, 
including those on metallicity, molecular abundance, dust properties, and the properties of the molecular gas.
For example, the high pressure in the interstellar medium as implied by
the superwind from the starburst probably helps to bind molecular clouds.
This increases the \twelveCO\ emissivity of the clouds and leads to an overestimation of
gas mass if the conversion factor derived for selfgravitationally bound clouds is applied
\citep{Bryant96, Oka98}.
Independent of these assumptions, dynamical mass sets a stringent constraint since the gas mass can not exceed it.
As we see in the next section (\S \ref{s.overall-kinematics}), the dynamical mass for the central region is estimated to be
\tilde$1.2\times 10^{10}$ \Msol\  for $r_{g} \leq 1.7$ kpc.

We adopt the total mass of neutral gas in the central region of
$M_{\rm gas}(r_g < 1.7 \mbox{ kpc}) \sim 2.9 \times 10^9 \Msol$ which is the logarithmic average of
the three mass estimates from \twelveCO(2--1), \CeighteenO(2--1), and dust emission.
For each of the three methods, we geometrically averaged the estimates given above,
except the one based on \twelveCO\ that exceeded the dynamical mass.
We assign our estimate a factor of 2 error in each direction, on the basis of the range of the three estimates.

\subsection{kinematics of the main gas disk  \label{s.overall-kinematics} }
The velocity field of the \twelveCO(2--1) emission in Fig. \ref{fig.naturalmap} shows
a spider pattern that is typical of gas disks in spiral galaxies.
The kinematical major axis is at a position angle of \tilde90\arcdeg\ in the vicinity of the double nucleus, 
and appears to be \tilde 70\arcdeg\ in the outer regions.
The outer position angle is consistent with those from previous spectroscopic observations
in CO(2--1) \citep{Aalto91a} and Br$\gamma$ \citep{Moorwood94}.
It also agrees with the isophotal fitting in the $K$-band by \citet{Rothberg04}, in which
the major axis has a nearly constant position angle of $\approx$60\arcdeg\
between the sky-plane radius of 2 kpc and 6 kpc.
The overall sense of rotation, i.e., east side being redshifted and west blueshifted, is the same as
that of \ion{H}{1} tidal tails \citep{English03}.
The apparent \tilde20\arcdeg\ shift in the kinematical major axis could be due to streaming motion of the gas 
or warp of the main gas disk.
The velocity field seen at this spatial scale (i.e., kpc scale) is relatively undistorted
for a merging galaxy.

The cold gas component appears to have kinematically merged (or relaxed) and formed a disk in the kpc scale.
We therefore fitted the \twelveCO\ velocity map in Fig. \ref{fig.naturalmap} using the AIPS task {\sf GAL}
to derive the kinematical parameters of the gas disk; the accompanying intensity map was used for data weighting.
The intensity weighting makes our parameters more appropriate for the inner
part ($r \lesssim 10''$) of the CO disk, 
reducing the effect of possible warp or non-circular motion in the outer disk.
The resulting dynamical center is between the two nuclei N and S.
In order to assess the robustness of the fitting results by altering the effect of beam smearing, 
we did the same fitting using the moment maps made with the uniform weighting,
which gave a better spatial resolution of $3\farcs3 \times 1\farcs3$ while retaining 87\% of
the flux in the natural weighting ($4\farcs8 \times 1\farcs9$) data.
The dynamical center, inclination, position angle of the major axis, and the systemic velocity agreed
within 0\farcs8, 8\arcdeg, 2\arcdeg, and 9 \kms, respectively.
The average parameters are
$\alpha_{\rm M}=10^{\rm h}27^{\rm m}51\fs25$, 
$\delta_{\rm M}=-43\arcdeg54\arcmin15\farcs7$ (J2000),
$i_{\rm M} = 45\arcdeg$,
$P.A._{\rm M} = 94\arcdeg$,
and
$V_{\rm sys, M} = 2791$ \kms, respectively, 
where the subscript M is for the main (or merged) disk.
One half of the abovementioned difference between the two fitting results gives 
each average value an estimate of uncertainty, though it is probably underestimated
because any systematic effect such as non-circular or non-coplanar motion of gas
could bias our fitting results.

The parameters we estimated are in reasonable agreement with those from previous observations.
Specifically, the inclination and position angle are consistent with those by \citet{Feast78} and others.
The systemic velocity is consistent with those by \citet{Casoli91}, \citet{Aalto91a}, and many others.
We use in this paper $V_{\rm sys} = 2775$ \kms\ taken from the literature. 
No more than 10\% error is introduced in the velocity related parameters derived in the following
--- such as dynamical mass, momentum, and energy --- by the slight difference (16 \kms) from our fitted value.
Lastly, but most notably, the dynamical center between the two nuclei is consistent with that measured
from \Halpha\ velocity field by \citet{Lipari00}.
The parameters we derived from the millimeter-wave CO line are not affected nor biased 
by dust obscuration, which is known to be large in the southern part of the main disk.
(For example, \citet{Kotilainen96} estimated $A_{V} \sim 10.7$ mag for line emission toward the S nucleus.)
As cautioned, the decoupled dynamical systems we suggest to exist around the double
nucleus (see below) may have affected the parameters to some extent.
We expect this effect to be small, because local distortion 
of the velocity field is not obvious around the double nucleus in the low resolution maps.

The dynamical mass within the radius of 10\arcsec\ $=$ 1.7 kpc of the main gas disk is estimated to be
$M_{\rm dyn}$(M, $r\leq 1.7$ kpc) $= (1.2\pm 0.4) \times 10^{10}$ \Msol\ from the rotation velocity
$V(r=10\arcsec)\sin i_{\rm M} = 125 \pm 20$ \kms\ and the inclination (45\arcdeg $\pm$ 4\arcdeg)
obtained from the kinematical fit.
(See Fig.  \ref{fig.high-res-pv} (c) for the rotation velocity.)
The fraction of gas in the dynamical mass is in the range of 0.1--0.5 for our estimate
of gas mass and the central value of the dynamical mass.

\section{Double Nucleus  \label{s.double} }
\subsection{gas distribution around the double nucleus}
The high-resolution map in Fig. \ref{fig.cor-2s} reveals compact gas peaks near or around
the nuclei N and S. 
The southern peak is more compact than the northern one, which  is more 
extended than the synthesized beam.
The two gas peaks are connected with a bridge of CO emission.
There are also arcs or partial spiral arms as well as secondary peaks that were already 
hinted in the lower resolution maps.
The CO distribution traces some of the conspicuous dust lanes in the optical, as shown
in Figure \ref{fig.coonhst} that compares the CO contours with the $I$-band HST image 
of \citet{Zepf99}.
Most notably, the following features correspond to dark lanes in the HST image:
the arc-like dust lane about 4\arcsec--8\arcsec\ east of the northern nucleus,
the gas peak about 5\arcsec\ east of the southern nucleus,
and the gas peak around the southern nucleus itself
as well as the westward tail emanating from the peak.
The `tail' feature of molecular gas actually trails the S nucleus, judging from the clockwise rotation of the
galaxy inferred from the spiral patterns in the HST image.

The \twelveCO(2--1) integrated intensity averaged within a 2\arcsec-diameter aperture
centered on each nucleus is 44 Jy  \kms\ \persquarearcsec\ for both of the nuclei.
This corresponds to a surface density of neutral gas $1.3\times 10^3$ \Msol\ \persquarepc\
on the sky plane if we use the same scale as our mass estimate in the previous section and
ignore the correction for missing flux. 
It also corresponds to a visual extinction of $A_{V} \sim 60$ mag for a uniform slab 
in front of the stars.
The actual extinction toward the nuclei is obviously much smaller than this because
at least near-IR images show both nuclei.
\citet{Kotilainen96} estimated $A_V \sim 2.4$ mag and $\sim 5.3$ mag 
for the continuum emission from N and S nucleus, respectively.
The reasons for these smaller extinctions are most likely 
that stars and gas are mixed,
that the gas and dust has a smaller scale height than the stars,
and that the molecular gas consists of clumps.

It is notable, however, that the S nucleus suffers more 
extinction than the N nucleus
despite the same gas surface densities inferred in the directions of the two nuclei.
The absorbing column density suggested from the actual extinction is a factor of 2.2 larger
for the S nucleus than for the N nucleus.
This discrepancy  suggests a geometrical effect.
The S nucleus is probably more deeply embedded in gas and dust or located behind the main gas disk.

The majority of \twelveCO(2--1) emission in the galactic center comes from the main gas disk,
despite the peaks around the two nuclei in the integrated intensity map.
The areas within 1\arcsec\ from the two nuclei have only 8 \% of the  \twelveCO(2--1) flux observed 
in the $r_{\rm g} \leq 10$\arcsec\ region,
which is the area within 10\arcsec\ $=$ 1.7 kpc from the midpoint of the double nucleus in the
plane of the main gas disk.
This fraction doubles, but is still small, if we apply a uniform correction for the flux
resolved out in the highest resolution map.

\subsection{gas kinematics and dynamical masses}

\subsubsection{disturbances in the main disk}
The high-resolution mean velocity map in Fig. \ref{fig.cor-2s} shows the same
overall rotation of the main gas disk as the lower resolution data. 
In addition  it reveals small scale disturbances in the disk.
The most notable of these outside the double nucleus is the noncircular motion
associated with the spiral feature about 4\arcsec--8\arcsec\ east of the N nucleus.
The merged gas disk is certainly disturbed at the few 100 pc scale.
However, we do not see in the main disk a double-peaked line profile that would have
suggested two distinct components overlapping in our line of sight.
It appears that most of the gas from the merger progenitors has coalesced in the 
main gas disk, except perhaps in close proximity to the two nuclei.

\subsubsection{possible mini-disks around the double nucleus \label{s.mini-disks}}
On, and very close to, each nucleus, the velocity map indicates a steep velocity gradient
in the east-west direction across the nucleus.
The steep velocity gradients are confirmed in the position-velocity (PV) diagrams shown
in Fig. \ref{fig.high-res-pv}.
The full velocity width within 1\arcsec\ of each nucleus is about 250 \kms\ and 310 \kms\
for the N and S nucleus, respectively.
Our data do not show a clear sign of velocity offsets, i.e., relative motion, between the main gas disk and
each nucleus.

We model the velocity gradient as rotation of gas around each nucleus.
The Keplerian dynamical mass within 1\arcsec\ (= 170 pc) of each nucleus is
$M_{\rm dyn}$(N, $r\leq 170$ pc) $= 6\times 10^{8} \sin^{-2} i_{\rm N}$ \Msol\ 
and 
$M_{\rm dyn}$(S, $r\leq 170$ pc) $=9\times 10^{8} \sin^{-2} i_{\rm S}$ \Msol\
in this model, where $i_{\rm N}$ and $i_{\rm S}$ are the inclinations of the
spin axes  with respect to our line of sight.
The uncertainty in these dynamical masses is probably as large as a factor of 2 
reflecting the uncertainty in the radii of the peak velocities.
For $i_{\rm N} \sim i_{\rm S} \sim 45\arcdeg$, the dynamical masses are approximately
$1\times 10^9$ \Msol\ and $2\times 10^9$ \Msol\ for N and S, respectively.

The dynamical mass for the N nucleus is consistent with (i.e., reasonably larger than) the optical
estimate, \tilde $10^8$ \Msol\ within 40 pc, based on the HST STIS spectra
across the nucleus \citep{Neff03, Lipari04}. 
The mass of neutral gas within 1\arcsec\ radius in the suggested disk is about
$1\times 10^8 \cos i_{\rm N}$ \Msol\ for the N disk (and $1\times 10^8 \cos i_{\rm S}$ \Msol\ for the S) 
if we scale the gas mass estimated in \S\ref{s.gas-mass-estimate} with \twelveCO(2--1) flux.
The gas-to-dynamical mass ratio therefore does not prohibit the mini-disk model.
The total mass of the two nuclei is at least 15 \% of the dynamical mass within a
10\arcsec\ (1.7 kpc) radius from the dynamical center of the merger.

In the mini-disk model, the suggested rotation axis of each nucleus is roughly parallel to that of
the main gas disk, and the sense of velocity gradient is about the same in the three disks
with eastward being redshifted and westward blueshifted.
This is consistent with the prograde-prograde merger that has been suggested for the galaxy.
In the merger configuration, the spin axes of  the progenitors and their orbital rotation axis are roughly parallel 
with each other and all rotation is in the same direction. 
Thus the dense gas disk to form around the core of each progenitor and the large gas disk to form from
merged disk gas will have spin vectors in roughly the same direction and orientation, as in NGC 3256, 
if the transfer of angular momentum is negligible between the spin and orbital motion of the mini disks.

Obviously one has to be careful about the interpretation of the velocity structure in the main gas disk
because of the unfortunate configuration that the two nuclei are aligned along the minor axis of the main disk.
This configuration 
makes it difficult to separate the rotation of the main disk and that around each nucleus
for a prograde-prograde system.
We therefore made a position-velocity diagram across the midpoint M of the N and S nuclei.
The PV diagram shown in Fig. \ref{fig.high-res-pv} (c) still shows the steep velocity gradient 
that we saw on the nuclei N and S.
There is, however, also a component with a shallower velocity gradient that almost linearly
rises to the velocity offset of \tilde 120 \kms\ at the offset of 10\arcsec.
We infer that this component with the shallow velocity gradient represents the main gas disk
and that the central component with a steep velocity gradient is the contamination from the N and S nuclei.
The contamination is quite possible  because the midpoint M is only 2\farcs5 from each nucleus 
while our spatial resolution in the north-south direction is 2\arcsec.

This interpretation is supported by the PV diagram through the two nuclei 
and their midpoint (Fig. \ref{fig.high-res-pv} (d)).
It suggests wider line widths at the two nuclei than at their midpoint.
Additional supporting evidence for the shallow rise of the rotation curve of the main disk is that
the isovelocity contours in the mean velocity map (Figs. \ref{fig.naturalmap} and \ref{fig.cor-2s}) are 
roughly parallel to the minor axis within \tilde5\arcsec\ from the dynamical center of the main disk. 
This indicates a near rigid-body rotation of the main disk in the region. 
It is consistent with the shallow velocity gradient in Fig. \ref{fig.high-res-pv} (c)  but not with 
a steeply rising rotation curve 
that approaches near its maximum velocity at a radius of \tilde1\arcsec.

Another cautionary remark is due for the outflow of ionized gas that \citet{Lipari00} suggested
to emanate from the N nucleus.
We regard it as unlikely to be the cause of the steep velocity gradient of molecular gas across the nucleus.
This is because the reported outflow is several kpc large and has
an outflow axis in the position angle of 150\arcdeg\ -- 160\arcdeg.
The flow axis is not aligned with the velocity gradient of the molecular gas but
roughly perpendicular to it.
Thus, although a starburst-driven wind can entrain molecular gas, 
the one reported from the optical observations would not  cause the observed CO velocity gradient.
Our estimate of the dynamical mass of the nuclei is therefore not affected by the outflow.
No report of an outflow has been made for the S nucleus so far, 
although, if it exists, it would not be easily detected through the large extinction.

A qualitative argument supports the suggested gas disks around the two nuclei, though
observational confirmation of them requires data at a higher resolution.
If near-IR and radio peaks N and S are indeed the remnant nuclei of the merger progenitors that
currently orbit in (or in and out of) the main gas disk, 
then collisional gas can follow each stellar nucleus to make a peak around it
only if  the gas is gravitationally bound to the nucleus. 
A massive remnant nucleus sinks toward the dynamical center because of dynamical friction.
The gas around the nucleus tend to be left behind if not bound to the nucleus,
because gas clouds are less massive than the nucleus and also because of hydrodynamical friction
between the gas around the nucleus and that in the main gas disk.  
For cold molecular gas bound to each nucleus,  the stable configuration is a rotationally supported disk
around the nucleus.
In this picture, some of the gas around each nucleus can be captured from
the main gas disk.

\subsubsection{comparison with other galaxies}
Small gas disks around merger nuclei separated by less than 1 kpc 
have been observed in Arp 220 \citep{Sakamoto99, Mundell01}.
Such a mini disk was also suggested in the center of M83 and attributed to a minor merger \citep{Sakamoto04}.
In Arp 220, the dynamical mass of each disk is at least $1$--$2 \times 10^{9}$ \Msol\ within a radius of 100 pc
and the two nuclei are separated by 0.3 kpc on the sky.

Another case of some similarity with NGC 3256 is
the luminous infrared galaxy NGC 6240, which is a merger with two nuclei separated by 750 pc on the sky.
\citet{Tecza00} found a steep velocity gradient across each nucleus in the stellar velocity field
measured with a CO absorption band, and inferred a pair of rotating {\it stellar} systems at the two nuclei
with a dynamical mass for each of  2--8 $\times 10^9$ \Msol\ within a radius of 200 pc.

The pairs of compact mass concentrations in the two major mergers Arp 220 and NGC 6240 
are comparable in size and mass with that in NGC 3256.
NGC 3256 may be a new member of the (yet) small group of merging galaxies that show
massive and kinematically-decoupled rotating disks in their central kpc.

\subsubsection{alternative models}
Lastly, there could be a totally different interpretation of our high-resolution observations.
Namely, one could assume that the putative nuclei N and S are not really remnant nuclei of the progenitors
but just bright star forming regions 
and that they are there because there are compact concentrations of gas for star formation \citep[c.f.,][]{Eckart01}.
This alternative would need a model to explain why there are two gas peaks in this galaxy and why
there appears to be large velocity gradients across the gas clumps.
We have not attempted such modeling.
If both N and S have a low-luminosity AGN as suggested by \citet{Neff03}, 
then they are most likely galactic nuclei.

We also did not explore the opposite model by \citet{Lipari00} that NGC 3256 has a third merger
nucleus at 6\arcsec\ east of the N nucleus.
The location is close to the CO spiral feature that we noted with its non-circular motion.
However, the non-circular motion appears to be associated with the spiral 
rather than localized on the suggested nucleus.
The line width at the position, at $-6$\arcsec\ in Fig. \ref{fig.high-res-pv} (a), is less than half of the ones
at the N and S nuclei. 
It is not significantly larger than the \tilde100 \kms\ line widths seen at other locations in the turbulent main gas disk
(see Figs. \ref{fig.high-res-pv} (b) and (c)).
The third nucleus, if exists, appears to play a minor role in gas dynamics.

\section{High-velocity gas \label{s.high-velocity-gas} }
\subsection{observational signatures}
We have discovered high velocity molecular gas in the galactic center.
This is most clearly seen in the  position-velocity diagram
in Figure \ref{fig.majpv} as the high velocity component at the offset of 0\arcsec.
The high velocity gas is seen in both blueshifted and redshifted velocities 
and is more prominent in the latter.
The \twelveCO\ PV diagram is along the major axis of the galaxy 
and is made by integrating the 6\arcsec\ wide area along the line of nodes.
The double nucleus is within this 6\arcsec-wide `slit' and is also at the offset of 0\arcsec.
The data cube used for this PV diagram has
4\farcs7 $\times$ 1\farcs8 (FWHM) and 10 \kms\ resolutions and has 92 \% of the
\twelveCO\ flux in the natural-weighting dataset.
In this dataset, the redshifted gas is detected up to the velocity of 3120 \kms, 
which is about 350 \kms\ from the systemic velocity of the galaxy.
The blueshifted gas at the galactic center is detected down to the velocity of \tilde2580 \kms,
or about 200 \kms\ from the systemic velocity,
though an arm-like feature about 15\arcsec\ west of the double nucleus has velocities down to about 2540 \kms.
We searched in the range of 2400 -- 3320 \kms\  for emission at even more extreme velocities, but
did not detect any at the abovementioned resolutions.
The redshifted component, however, appears to have a faint tail up to the velocity 
of \tilde3200 \kms\ in a lower-resolution dataset as we see below.
We did not detect the high velocity gas in \thirteenCO\ and \CeighteenO.

The high velocity gas is located {\it between} the two nuclei.
This is seen in Figure \ref{fig.blue-red-gas} that shows \twelveCO(2--1) maps made by averaging 
the velocities of the redshifted and blueshifted high velocity gas. 
In both maps, the high velocity gas is not peaked on either of the two nuclei but between them.
The peak signal-to-noise ratio of the high velocity gas in the maps is 12.8 and 4.4 for the 
redshifted and blueshifted gas, respectively. 
Any error in continuum subtraction can hardly affect this emission, because the maximum
flux density of the continuum is less than a third of the peak flux density of the fainter,
blueshifted high velocity gas. 
The peak position of the redshifted high velocity gas is
$\alpha =10^{\rm h}27^{\rm m}51\fs22$ and
$\delta  =-43\arcdeg54\arcmin16\farcs3$ (J2000)
according to a Gaussian fit.  
The emission has a deconvolved size of $4\farcs6 \times 2\farcs5$ (P.A.=161\arcdeg).
The  emission is too weak, and possibly too extended, to be detected in our uniform weighing maps
used in the previous section.

The high velocity gas appears almost at the same position in the redshifted and blueshifted velocities.
Although the blueshifted emission is too weak to precisely determine its centroid position, 
the peak positions in the two maps in Figure \ref{fig.blue-red-gas} agree within 1\farcs3.
In addition, we did not find a systematic velocity gradient within the redshifted high velocity gas.
To check this, we made a mean velocity map of the high velocity gas to see 
velocity gradient in any particular direction,
and also inspected channel maps to see if there was a systematic shift of the emission centroid.
We did not detect a sign of velocity gradient in either of them.
This can constrain the parameters in models for the high velocity gas
(e.g., rotation curve for a rotation model and outflow geometry for an outflow model).

The integrated \twelveCO(2--1) flux of the high velocity gas is 3 \% of
the total \twelveCO(2--1) flux in the 5\arcsec-diameter area at the galactic center.
This is measured from the spectrum in Figure \ref{fig.hvgspec}. 
The spectrum is extracted at the midpoint of the double nucleus 
from our natural weighted data convolved to the 5\arcsec\ resolution. 
The emission line spans from 2530 \kms\ to 3200 \kms\ when we use the first null
on each side to determine the line edge.
The full width at zero intensity (FWZI) of the emission is thus 670 \kms.
The CO flux in the red component above 2990 \kms\ is 23 Jy \kms\
in the central 5\arcsec, 
that in the blue component below 2610 \kms\ is 5 Jy \kms, 
and the total flux in the line is 936 Jy \kms.

\subsection{molecular outflow model}
\subsubsection{supporting evidence for outflow}
We suggest that the high velocity gas is a bipolar outflow of molecular gas from the main gas disk.
This is mainly because the galaxy is known to have a superwind extending several kpc
out of the central starburst \citep{Scarrott96, Moran99, Heckman00, Lipari00} and
the flow of hot gas can entrain cold molecular gas.
It is also because alternative explanations of the high velocity gas seem less likely as we see below.
The location of the high velocity gas makes the main gas disk the most likely place from which
the molecular wind emanates.

Among the existing observations of the superwind, \citet{Heckman00} detected 
toward the central few arcsecond of NGC 3256
a doublet of \ion{Na}{1} D absorption line blueshifted by about 300 \kms\
with respect to the systemic velocity of the galaxy.
In our CO observations, the maximum velocity offsets are $+420$ and $-240$ \kms\ with
respect to the systemic velocity, according to the FWZI line width.
The velocity offsets of the \ion{Na}{1} and CO lines are in good agreement
in magnitude.  
Both lines trace neutral gas;  the former traces atomic gas and the latter molecular gas.
Thus the agreement supports the outflow model of the high velocity molecular gas.

\subsubsection{alternatives}
Before going further into the outflow model, we briefly check its alternatives.
First, rotation can cause high velocity.
It would have to be the one around the dynamical center of the merger,
where the high velocity gas is.
The dynamical mass needed for the high velocity would be
at least $3\times 10^{9}$ \Msol\ within a radius of 110 pc (=0\farcs65)
if we use the CO line width for velocity and
the nominal offset between the blueshifted and redshifted CO peaks for size.
The dynamical mass is larger than those of the two nuclei (\S \ref{s.mini-disks}) despite the
lack of a $K$-band peak between them.
In addition, the large mass does not fit the rotation curve that we inferred 
in \S \ref{s.mini-disks} to be nearly linearly rising to the radius of 10\arcsec.
The dynamical mass can be smaller and consistent with the observations
if the extent of the high velocity gas is smaller, but a smaller gas disk means
a higher mass density and a higher CO intensity in it.
For example, a disk with a 10 pc radius
would require the dynamical mass, mass density, and brightness temperature of
$3\times 10^{8}$ \Msol, $5\times 10^{4}$ \Msol\ \percubicpc, and 100 K,
respectively.
An AGN may realize these conditions, but none has been found between the
two nuclei. 
Thus we regard the rotation model less likely than the outflow model.

Other alternatives include locally large velocity dispersion, gas infall from out of the main disk, 
and an outflow driven by an AGN jet. 
The velocity dispersion does not seem to have a local source of turbulence, 
unless there is a special mechanism related to the merger gas dynamics.
The gas infall from both (i.e., approaching and receding) sides toward the dynamical center seems 
too much a coincidence. 
No AGN jet has been seen in the region.
Note that all the alternative models except the jet-driven outflow can not explain the blueshifted
\ion{Na}{1} absorption, though the CO and \ion{Na}{1} features could be of different origins.

\subsubsection{outflow parameters}
In the outflow model, we can estimate the following parameters from our observations.

{\it Velocity: ---} If the outflow is well collimated along the rotation axis of the main gas disk then the
maximum flow velocity corrected for inclination would be 
600 \kms\ for the receding flow and 350 \kms\ for the approaching flow.
However the outflow may have a wide opening angle.
The inclination correction in that
case would be smaller or even negligible.

{\it Mass: ---} The mass of molecular gas involved in the high-velocity flow is computed to be
$1.4\times 10^7$ \Msol\ and $0.3\times 10^7$ \Msol\ for the receding and approaching gas, respectively, 
from the \twelveCO(2--1) flux in the velocity ranges where the high velocity gas is distinguishable
from the disk gas. 
These estimates are based on the \twelveCO(2--1)-to-gas mass scaling factor that we adopted for the
galactic center (\S\ref{s.gas-mass-estimate}). 
Their uncertainties are larger than that for the disk gas because we do not know 
whether and how the \twelveCO(2--1) emissivity changes in the outflow gas.
The contribution from atomic gas, which was a factor of 0.3 in the gas mass estimated in \S\ref{s.gas-mass-estimate}, 
is excluded from the above numbers because atomic-to-molecular mass ratio in the outflow is likely different from
 that in the disk gas.
The numbers given here suggest that 
{\it molecular gas of the order of $10^7$ \Msol\ is being ejected from the
central kiloparsec at velocities $\gtrsim 200$ \kms\ in the molecular outflow.}
There is likely a larger amount of molecular gas in the outflow at smaller velocities.

{\it Momentum and Kinetic Energy: ---}
The momentum and kinetic energy of the molecular outflow are estimated using the same mass
scaling and integration on the spectrum in Fig. \ref{fig.hvgspec}.
The red and blue high velocity gas have momentum of $8\times 10^{42}$ kg \ms\ and $1\times 10^{42}$ kg \ms,
respectively, in the direction of our line of sight.
The kinetic energy of the redshifted high-velocity gas is $1\times 10^{48}$ J, without correction for
the inclination of the flow, while the blueshifted gas has one tenth of that energy.
The unknown contribution of outflow gas slower than 200 \kms\ becomes smaller for the
momentum and kinetic energy because of the $(\Delta v)^{1}$ and $(\Delta v)^{2}$ weights
used for these moments.

{\it Timescale : ---}
We estimate the characteristic timescale for the redshifted outflow with more than 200 \kms\ velocity offset
to be \tilde 2 Myr from the deconvolved size of the high velocity emission and the outflow velocity.
This is not the age of the outflow, which should be comparable to the age of the starburst and is 
probably an order of magnitude larger.
The timescale is rather the crossing time, for the high-velocity gas, of the $\lesssim 500$ pc region
where the high velocity emission is detectable.
For symmetry, we assume the same timescale for the
blueshifted high velocity gas.

{\it Fluxes and Mechanical Luminosity: ---}
We can estimate the mass flux, momentum flux, and mechanical luminosity of the high velocity molecular
outflow using the timescale and the parameters estimated above.
The mass flux of molecular gas in the high velocity outflow is \tilde9 \Msol\ \peryr, 
the momentum flux injected to the outflow   
is  \tilde$1\times 10^{29}$ N,
and the kinetic luminosity of the outflow gas is \tilde$2\times 10^{34}$ W.
These values are for the high velocity molecular component.
Contributions from atomic and ionized gas and those from slower velocity flow
increase these numbers. 
For comparison, the average rates of
$17^{+20}_{-9}$ \Msol\ \peryr, $10^{29.1 \pm0.5}$ N, and $10^{34.6\pm0.6}$ W
have been reported for a sample of infrared luminous galaxies ($10^{11.36 \pm 0.4}$ \Lsol) 
from \ion{Na}{1} observations \citep{Rupke05}.

The uncertainties in the parameters derived here are worth summarizing.
The parameters involving gas mass inherit the uncertainties of the mass estimate, with
an additional uncertainty from the possible variation of gas properties between the
disk and the outflow.
Those involving velocity have uncertainty due to the unknown outflow geometry.
We did not apply an inclination correction, in effect assuming a wide opening angle of the flow.
Still, velocity-related parameters are estimated from velocity components along our line of sight.
If we apply the correction for a well-collimated outflow 45\arcdeg\ inclined to our line of sight,
then momentum and kinetic energy increase by a factor of 1.4 and 2, respectively.
The momentum flux and the kinetic luminosity also increase by a factor of 1.4 and 2, respectively,
since the crossing time does not change with the correction.
As noted, the parameters do not include the contribution of the outflow gas whose low 
line-of-sight velocity does not distinguish the gas from the disk gas.

\subsubsection{energy and gas consumption budgets}
The energy of the outflow can be supplied by a portion of the starburst.
If we equally divide the far-IR luminosity of the galaxy among each of the two nuclei and the main disk,
then the mechanical luminosity released from the starburst in the main disk is \tilde $3\times10^{35}$ W 
assuming a 10 Myr-old continuous starburst that has
the Salpeter initial mass function (IMF) in the mass range of 1--100 \Msol\  \citep{Leitherer99}.
Thus the molecular outflow is energetically possible with a wide margin.
It can be caused by passing $\lesssim 10$ percent of the mechanical energy
from the disk starburst to molecular gas; the rest can be used to drive the superwind of hot gas.

The star formation rate needed to generate the starburst luminosity is \tilde10 \Msol\ \peryr\ under the
same assumptions on the starburst.
The rates are therefore of the same order of magnitude
for the consumption of molecular gas by star formation and 
for the dispersal of molecular gas by the outflow.
This comparison, however, should be viewed with caution because the star formation
rate estimated from luminosity, as well as the gas outflow rate, has a large uncertainty.
The former rate can change by a factor of 10 with different assumptions about the IMF, because high mass stars
generate most luminosity while low mass stars usually determine most of the gas consumption.

Perhaps a more robust number to describe the significance of the molecular outflow is
the depletion time of molecular gas solely by the outflow.
It is estimated to be 70 ($=2/0.03$) Myr from the fraction of high velocity gas in the central 5\arcsec, 0.03, and
the time scale of the outflow, 2 Myr,  under the simplistic assumption 
that the high velocity gas does not return\footnote{
We can not easily assess whether the molecular gas escapes the galaxy or not,
without information about the current height of molecular gas above the main disk, 
the shape of the gravitational potential, and structure of the superwind.
If the gas returns, then the timescale is the one during which all the gas experiences the out-of-plane trip 
at least once. The molecular gas may well be dissociated and ionized in the hot wind during the trip.
 }.
The timescale is comparable to the typical timescale of a starburst, 
again implying the significant role of the molecular outflow in the starburst.
This timescale of gas depletion or dispersal by the outflow is estimated
without explicitly using the CO-to-\HH\ conversion factor.
Hence the uncertainty in the depletion time due to the conversion factor is only indirect,
through any variation of the scaling factor between the outflow and disk gas.
We caution that the timescale is derived to measure the significance of the molecular outflow rather than
to predict the evolution of the system, for which one needs to consider such factors as the
time evolution of star formation, that of the outflow, and further gas accretion within the galaxy.

To summarize, our observations of the molecular outflow agree with 
the optical absorption-line surveys toward starburst galaxies \citep{Heckman00, Rupke05} in that
a starburst-driven superwind can be as important as the star formation itself in the
gas consumption budget of a luminous infrared galaxy.
\citet{Rupke05} obtained a median value of 0.33 for the mass entrainment efficiency (i.e.,
the rate of mass loss due to superwind normalized by the star formation rate) 
in luminous infrared galaxies, to which NGC 3256 belongs.
The efficiency for the molecular outflow in NGC 3256 is \tilde1 if a third of the starburst
in the galaxy contributes to the observed outflow, or \tilde0.3 if the entire starburst contributes.
Thus the entrainment efficiency for molecular gas in NGC 3256 agrees, in the order of magnitude,
with the median efficiency estimated for atomic gas.

\subsubsection{geometry}
We finally comment on the spatial configuration of the starburst in the galactic center.
The spatial distribution of massive star formation is peaked toward the N nucleus
as seen in [\ion{Fe}{2}] and Br$\gamma$ \citep{Moorwood94}, Pa$\alpha$ \citep{Alonso-Herrero02}, 
H$\alpha$ and diffuse X-ray in the 0.3--10 keV range \citep{Lira02}.
The S nucleus also shows a peak in Br$\gamma$, though its flux corrected for extinction
is estimated to be 4 times smaller than that of the N nucleus \citep{Kotilainen96}.
In contrast, the region between the two nuclei does not have a peak in these tracers of massive star formation,
even though there are \Htwo\ regions and super star clusters spreading 
between and around the two nuclei \citep{Alonso-Herrero02}.
It is therefore somewhat puzzling to find a molecular outflow between the two nuclei and not from each nucleus, 
in particular from the nucleus N.
Unfortunately the optical observations that indicated the presence of a superwind do not tell us 
the exact geometry of the wind and whether there are multiple superwinds in the region.

Clues to solve the puzzle may be that the base of the superwind likely has a size of 1 kpc or larger
and that the two nuclei as well as the star forming clusters in the region 
have been moving with respect to the dynamical center of the system.
\citet{Heckman90} measured the pressure profile in the center of NGC 3256.
They found that the central 1--2 kpc diameter region has very high pressure (\tilde$10^{-9.5}$ Pa) 
with a shallow decline outwards, and that the pressure gradient steepens outside the region.
As they interpreted the profile, this is expected for a superwind with a base diameter of 1--2 kpc,
within which static thermal pressure dominates \citep{Chevalier85,Tomisaka86}.
The starburst region, i.e., the region producing the high pressure in this case, thus encompasses 
the two nuclei. 
The two nuclei are about 500 pc from the dynamical center of the system and probably have
a velocity of about 100 \kms\ with respect to the dynamical center.
Thus their crossing time across the central kpc is about 10 Myr. 
It is comparable to the typical timescale of a superwind.  
The two dominant energy sources thus move while a superwind develops, effectively making the
region of energy injection larger than that of the two nuclei.
Once a kiloparsec-size region of hot gas is formed by the moving energy sources, 
its evolution as a superwind is
governed by the large scale structure of the ISM in which the hot gas is embedded.
It is therefore conceivable that a main superwind centered around
the dynamical center of the system blows out in the direction perpendicular to the main gas disk.
This solution to the puzzle, however, seems to have a much room for improvement.

\section{Molecular Gas Properties \label{s.gas-properties}}  
\subsection{temperature and filling factor \label{s.gas_temp}}
Molecular gas in the center of NGC 3256 is warm. 
This can be seen in the map of peak brightness temperature shown in Fig. \ref{fig.cor-2s}.
The peak temperature in the map, in excess of 10 K, is high considering that it is averaged 
over the 170 pc $\times$ 340 pc beam.
The map is made by searching for the maximum brightness temperature in the velocity direction
of the data cube at each sky position.
The peak brightness temperature gives a lower limit on the temperature 
of molecular gas at each position, except for low intensity regions dominated by noise.
It is a lower limit because of beam dilution --- the filling factor of gas in the beam can be much smaller than unity ---
and also because the emission may be optically thin.
In particular, the map tends to give lower temperatures to regions with large line widths
if the gas physical temperature and surface density are the same among regions.
Despite these difficulties in reading the map, the map is indicative of higher gas temperature around the N nucleus
and  the tail region extending west from the S nucleus.
As we noted, the N nucleus is the most active star forming region in the galactic center.
The tail region corresponds to the tail feature seen in diffuse X ray emission \citep[Fig. 12]{Lira02}.
The small temperature depression on top of the peak at the N nucleus and unremarkable temperature at the S nucleus
are at least partly due to the dilution caused by the large line widths there.

To put the beam diluted brightness temperature into perspective, we note that the peak value of 13 K in NGC 3256 
\citep[$L_{8-1000 \micron}=10^{11.56}$ \Lsol ;][]{Sanders03} 
is a factor of 3 lower than that in the ultraluminous infrared galaxy Arp 220 ($10^{12.21}$ \Lsol) 
observed with the same line at almost the same linear resolution \citep{Sakamoto99}.
It is also four times higher than the peak temperature of 3 K 
obtained in the starburst nucleus of M83 ($10^{10.10}$ \Lsol)
by convolving the CO(2--1) observations of \citet{Sakamoto04}  to match in linear resolution.
All three galaxies are moderately inclined with respect to our line of sight 
($i \sim$ 25\arcdeg\ -- 45\arcdeg)
and hence are expected to be similar in the degree of dilution due to galactic rotation. 
This comparison implies that molecular gas in galaxies with higher luminosity tends to be 
warmer or to have higher area filling factors or both when compared at 200 pc scale.

\subsection{gas cloud properties and line excitation}
\subsubsection{line ratio observations}
NGC 3256 has a high \twelveCO/\thirteenCO\ intensity ratio. 
Figure \ref{fig.cospectra} compares the spectra of the three CO lines 
obtained with a simulated 30\arcsec\ beam (FWHM) centered on the
midpoint of the double nucleus.
The flux densities integrated from 2590 \kms\ to 2970 \kms\ are
$4.62\times 10^{3}$, $1.37\times 10^{2}$, and $26.1$ in unit of Jy \kms.
The errors of these integrated intensities are \tilde10\% due to flux calibration
and $\pm2.6$ Jy \kms\ due to noise directly estimated from the maps.
The flux calibration error, however, cancels out in a line ratio 
from our simultaneous observations.
Thus the error in the line ratio is dominated by noise.
We obtain integrated temperature ratios of 
$R_{12/13} (2-1) \equiv I(\twelveCO\, 2-1) /  I(\thirteenCO\, 2-1) =  30.9 \pm 0.6$,
$R_{12/18} (2-1) \equiv I(\twelveCO\, 2-1) /  I(\CeighteenO\, 2-1) = 161 \pm 15$,
and
$R_{13/18} (2-1) \equiv I(\thirteenCO\, 2-1) / I(\CeighteenO\, 2-1) = 5.2 \pm 0.5$,
where $I(\mbox{line}) \equiv \int \! T_{\rm b}(\mbox{line})dv$.
Our $R_{12/13}$(2--1) ratio agrees with those from previous single-dish
observations (see Fig. \ref{fig.coratio}).
The $R_{12/13}$(2--1) ratio averaged over 5 kpc in NGC 3256 is about three times higher than the ones observed
in the central kpc of our Galaxy, $10\pm 1$ \citep{Sawada01}, central 300 pc of
the starburst galaxies IC 342 and NGC 253, $\sim 9$ \citep[and Fig. \ref{fig.coratio}]{Meier00},
and in the central 500 pc of M82, \tilde12 \citep{Mao00}.
It is also higher than the $R_{12/13}$(1--0) values of  $\approx 11$ observed in the centers of nearby
spiral galaxies  \citep{Sage91,Aalto91b, Paglione01}.
On the other hand, the high $R_{12/13}$(2--1) ratio is within the range of values observed
among luminous mergers, $\gtrsim 20$ \citep[e.g.,][]{Casoli92b,Aalto95,Glenn01}.

  The intensity ratios are not uniform in the galaxy.
 Our data indicate a trend that the $R_{12/13}$(2--1) ratio increases with
 the beam size (Fig. \ref{fig.coratio}), suggesting higher ratios in the outer parts of the galaxy.
 The same trend is seen in $R_{12/18}$(2--1) with less significance.
 The $R_{13/18}$(2--1) ratio has too large uncertainty to detect its radial variation.
 The two nuclei have the lowest $R_{12/13}$(2--1) value of 17 in our measurements. 
 One can infer that the ratio is locally low at the two nuclei 
 by looking at the integrated intensity maps in Fig. \ref{fig.naturalmap} where 
 \thirteenCO(2--1) shows peaks at the nuclei while \twelveCO(2--1) does not.
The radial gradient of the  \twelveCO/\thirteenCO\ intensity ratio in the J=2--1 transition
is opposite to that in the J=1--0 emission.
The $R_{12/13}$(1--0) ratio at the center of the galaxy is high,  26--40 \citep{Aalto91a,Casoli91,Becker91,Garay93}, 
but it decreases  to 11 at  43\arcsec\ NE of the galactic center \citep{Aalto95}.

\subsubsection{gas model for the nucleus and the main disk}
The  two-component model of molecular gas suggested by \citet{Aalto95}
seems to work for our higher central $R_{12/13}$(2--1) value than in the centers of 
less luminous galaxies.
In the model, 
molecular gas in a starburst nucleus has
a significant amount of warm and turbulent envelope gas with moderate opacity 
($\tau$[\twelveCO(1--0)] $\sim 1$ and $\tau$[\thirteenCO(1--0)] $\ll 1$) 
around cloud cores of high optical depths
($\tau$[\twelveCO(1--0)] $\gg 1$ and $\tau$[\thirteenCO(1--0)] $\sim 1$).
The large contribution (i.e., area filling factor) of the warm envelope gas makes the intensity-weighted optical depth
of \twelveCO\ emission moderate (\tilde1) and makes the $R_{12/13}$(1--0) ratio higher
than in core-dominated quiescent gas.
The warm gas with high area-filling factor noted in \S\ref{s.gas_temp}
fits the profile of the envelope gas.
The dense and high-opacity cores, on the other hand, are supported 
by the \thirteenCO(2--1)/\thirteenCO(1--0) ratio close to unity, 1.3, in the center of NGC 3256 \citep{Casoli92b}
and by the observations of dense-gas tracers such as HCN \citep{Casoli92a}.
For the $R_{12/13}$ ratio in the J=2--1 transition,  
we note that the transition tends to have a higher opacity than the J=1--0 one.
Thus we expect
higher opacities in the envelope gas or, equivalently, a larger filling factor of optically thick cores. 
This and the presence of envelope gas make 
the $R_{12/13}$(2--1) ratio high at the galactic center but smaller than $R_{12/13}$(1--0), as observed.

The model needs to be slightly modified in the outer disk to accommodate the opposite radial trends
in $R_{12/13}$(1--0) and $R_{12/13}$(2--1).
For the outer disk, the same argument as above 
would lead to a lower $R_{12/13}$(2--1) value in the disk than in the nucleus.
It is because, according to the model, the disk emission is dominated by cloud cores.
Such a radial trend of the J=2--1 ratio contradicts our observations.
One way to alleviate this is to assume that the CO excitation is subthermal in the disk cloud cores
and that the cores have lower opacities in the J=2--1 transition than they are expected to have from their J=1--0
opacities if the cores were in LTE.
In other words, many cloud cores in the merger disk may look like low-opacity envelopes in the J=2--1 lines. 
As a result, in the J=2--1 lines, the bulk of emission in the galaxy has moderate (in \twelveCO) and 
moderate to low (in isotopes) optical depths throughout, with opacities increasing toward the center.
The radial gradient of the $R_{12/18}$(2--1) ratio and the insignificant gradient of the $R_{18/13}$(2--1) ratio
are consistent with this.
The subthermal excitation in the disk could be due to lower gas density there than in the nucleus,
large velocity dispersion in the merged gas disk, and 
lower fractional abundance of CO with respect to \HH.
The higher [\twelveCO]/[\thirteenCO] abundance ratio expected from the fresh supply of 
low-\thirteenC\ gas from peripheries of the merger progenitors  \citep{Casoli92b, Henkel93}
would also increase the $R_{12/13}$(2--1) ratio in low opacity clouds.
Our conceptual model for the radial gradient of gas properties in NGC 3256
can be better constrained with more data. In particular, additional high-resolution 
data in other CO transitions, such as CO(1--0) or CO(3--2) would be most helpful
to estimate the excitation conditions.

Our observations in terms of radial gradient of the $R_{12/13}$(2--1) ratio
are consistent with those  in the Antennae galaxy.
\citet{Zhu03} found that $R_{12/13}$(2--1) was about twice as high in the overlap region of the merger
than in the centers of the colliding galaxies. 
Their galactic center ratios, 
$13\pm3$ and $16\pm4$ for each galaxy, are marginally higher than the ones we quoted
for local starburst nuclei.
The higher $R_{12/13}$(2--1) ratio in the overlap region, however, may be because
the region has a starburst and resembles merger nuclei.
\citet{Aalto95} mentioned a decreasing $R_{12/13}$(2--1) gradient from the nucleus outwards in 
the polar ring galaxy NGC 660 
and an increasing gradient in the merging galaxy NGC 2146.
Gas properties and excitation conditions of CO may be diverse among mergers
in their outer disks.

\section{Discussion \label{s.discussion} }
Our observations have provided a wealth of new information on the molecular interstellar medium
in the center of NGC 3256.
We can place them in the general context of the galaxy merger
and that of merger-induced starburst. 
They also point areas for further studies.

The main gas disk is what is expected from the dissipative nature of the interstellar medium
and what has been predicted from numerical simulations 
\citep[e.g.,][]{Mihos96, Barnes96}. 
In the simulations, gas from the disks of progenitor galaxies collides and eventually coalesces 
into a single disk rotating around the dynamical center of the merger, 
with a strong concentration toward the center.
Such a merged gas disk has been observed in a number of luminous and ultraluminous
merging galaxies \citep[e.g.,][]{Scoville97, Downes98, Bryant99, Tacconi99}.
In line with those models and observations, the main gas disk of NGC 3256 has the majority of 
molecular gas in the system
and the overall gas distribution peaks toward the center of the merger.
The main gas disk is certainly supported by rotation, with  the center of rotation
located between the two nuclei.
Presumably reflecting the violent past of the system, however, 
the disk shows signs of disturbances in the forms of short spiral arms and non-circular motions.

Our observations reinforce the notion that  properties of molecular gas in this merger
as well as other luminous mergers
are different from those in our Galaxy in the sense that the former
contains more of warm, tenuous (i.e., high area-filling factor), and turbulent molecular gas \citep{Aalto95, Scoville97, Downes98}.
This must be due to the galaxy merging and to the starburst in the center of the merger.
Our observations also suggest that gas properties change in spatial scales from
over a kiloparsec to as small as a few 100 pc, which is our highest resolution.
Examples of such spatial variation include 
the highest brightness temperature gas associated with the 
N nucleus that hosts the most active star formation in the galaxy, 
the local minima  of \twelveCO(2--1)/\thirteenCO(2--1) ratio on each of the two nuclei,
and
the global radial gradient of the intensity ratio in the main gas disk.
While these gas properties are probably the results of the merger and the starburst, 
the gas properties are also expected to influence the star formation from molecular gas.
Therefore the study of gas properties at high resolutions helps to 
model the evolution of a starburst.
We noted a trend of higher CO brightness temperature in more luminous galaxies.
Statistical studies to probe trends like this and trends in line ratios will help characterize 
the molecular gas in mergers and starbursts.

The {\it molecular} outflow from a merger has little observations
in luminous or ultraluminous infrared galaxies of $\log(L_{\rm IR}/\Lsol) \geq 11.5$,
though surveys of atomic and ionized gas have shown that
most luminous and ultraluminous infrared galaxies have a superwind 
\citep{Heckman90, Heckman00, Martin05, Rupke05}.
The recent detection by \citet{Takano05} of high velocity gas in NH$_3$ absorption toward Arp 220  
may be one of the rare observations of molecular outflow from a luminous
infrared merger, although other interpretations are possible for the absorbing gas.
In the model of superwind,
an intense starburst deposits a large amount of kinetic energy from stellar winds and supernovae
into the interstellar medium. The energy is thermalized in the ISM to generate hot gas of $10^{6}$--$10^{7}$ K 
that can not be contained in the galactic potential and blows out of the galactic disk. 
Although molecular gas can not be hot enough to blow out by itself, it is expected to be
entrained by the wind of hot gas.
There are examples of such entrainment of molecular gas in nearby starburst galaxies of lower luminosities, 
e.g., in M82 \citep{Nakai87, Seaquist01, Walter02}. 
It is therefore not surprising to detect a molecular outflow from such a luminous starburst as NGC 3256
in sensitive observations. 
The outflow parameters estimated in NGC 3256, i.e., the maximum outflow velocity,
mass flux, momentum flux, and mechanical luminosity, are in order-of-magnitude agreement
with those obtained from the survey of atomic Na absorption in luminous and ultraluminous
infrared galaxies by \citet{Rupke05}.
In particular, we reiterate the importance of gas outflow in the gas budget of the galactic center;
molecular outflow can be nearly as significant as star formation in gas consumption.
It is almost certain that we detected only a portion of the molecular outflow in NGC 3256, limited by our sensitivity.
Observations at higher sensitivity and resolution would shed more light
on the mechanism for molecular gas entrainment and the effects of the evacuation of molecular gas
on the starburst evolution.
Such observations will also help testing the alternative interpretations of the high velocity gas.

Regarding the double nucleus, we suggest that the gas peaks around and the steep velocity gradient
across each nucleus are most likely due to a mini disk rotating around the nucleus, 
although the mini disks certainly merit  further observations for confirmation.
There is a striking similarity among NGC 3256, Arp 220, and NGC 6240
in terms of their twin (possible) remnant nuclei of a few 100 pc sizes and \tilde $10^{9}$ \Msol\ masses
within the projected separation of 1 kpc.
In the earlier phase of galaxy collision, numerical simulations
suggest that a sizable fraction of gas in each galaxy can be funneled to the nucleus of each galaxy,
forming a massive rotating core of gas and that of stars through starburst \citep{Noguchi88, Barnes96, Mihos96}.
Observations of interacting galaxies find such starbursting gas concentrations in many, though not all, galaxies
\citep{Aalto97,Wilson00,Yun01,Kaufman02,Evans02}.
As the merger proceeds, stellar systems will merge from outside, leaving nuclear stellar cores until the final stages
\citep{Funato92,Boylan-Kolchin04}. 
Gas from the merger progenitors will also coalesce first from outside to form a large merged disk,
which is most likely the main gas disk in NGC 3256.
In the mean time, the nuclear gas trapped deep inside the gravitational potential
of each progenitor likely remains within the nucleus.
The pairs of massive nuclei in the three merging galaxies are plausibly the remnants of such massive cores of 
stars and gas. (Gas was presumably striped from the stellar cores in NGC 6240.)
The massive remnant nuclei should strongly perturb the larger, merged gas disk as they
sink toward the dynamical center of the system owing to dynamical friction.
Recently \citet{Escala05} suggested from numerical simulations of a binary massive black hole in a merger 
that cold gas in the central merged disk can be  a significant source of dynamical friction 
and that the gas disk is affected in return forming a gap or spirals or both around the binary.
The situation may be similar in NGC 3256.
The gas concentration around each nucleus should eventually
contact the gas around the other nucleus or in the main disk, loses angular momentum, and probably
falls toward the dynamical center of the system to become a part of the main gas disk.
The complicated dynamics of gas, stars, and presumably massive black holes,
remains to be studied along with the evolution of starburst in the perturbed region.
Our observations show that NGC 3256, along with Arp 220, is one of the
best targets for such studies.

\section{Conclusions \label{s.conclusions}}
We have made the first interferometric observations of molecular gas and dust  in the center of the
southern luminous merger NGC 3256, improving the spatial resolution
over previous single-dish observations by an order of magnitude.
Our main findings are the following:

\begin{enumerate}
\item 
There is a large disk of molecular gas ($r > 3$ kpc) with a strong concentration toward the double nucleus.
The velocity field of the disk suggests rotation around a point between the two nuclei.
The mass of neutral gas in the central 20\arcsec\ ($r \leq 1.7$ kpc) is estimated to be 
\tilde $3 \times 10^9 \Msol$.
Our observations suggest that the molecular gas in the merger system has mostly merged and settled 
into this main disk at the radii of 1--5 kpc.

\item 
In the central kiloparsec, our highest resolution ($1\arcsec \times 2\arcsec =$ 170 pc $\times$ 340 pc) image
shows that molecular gas peaks near or at each of the two nuclei.
There is also a steep velocity gradient across each nucleus, suggestive of a small gas disk
rotating around the nucleus.
The dynamical mass of each nucleus  is estimated from the gas velocity to be \tilde$10^{9}$ \Msol\
within a radius of 1\arcsec\ (= 170 pc).
The suggested configuration --- two mini disks around the binary nuclei 
along with a larger main disk --- is similar to that in Arp 220.

\item 
We discovered high velocity molecular gas of \tilde$ 10^{7}$ \Msol\ at the galactic center between the two nuclei.
It is up to about 400 \kms\ from the systemic velocity of the galaxy.
The terminal velocity is about two times larger than that due to rotation of the main gas disk.
We have modeled the high velocity gas as a molecular outflow, entrained by the superwind that
the starburst in the central kiloparsec is known to have created.
Outflow parameters, including the outflow rate of \tilde10 \Msol\ \peryr, are estimated from the data.
They suggest that the dispersal of molecular gas from the starburst region can be as significant
as the gas consumption by star formation in the gas budget.

\item  
Our simultaneous observations of the \twelveCO(2--1), \thirteenCO(2--1), and \CeighteenO(2--1) emission
provided accurate measurements of  intensity ratio 
among those lines. 
The \twelveCO/\thirteenCO\ intensity ratio in the $J$=2--1 transition is higher at the nucleus 
than in the centers of our Galaxy and nearby starburst galaxies.
Our observations are consistent with the model that the intense starburst in the merger nucleus  
is taking place in warm turbulent gas with a high filling factor and moderate \twelveCO\ opacity. 
Positive radial gradients of \twelveCO/\thirteenCO\ and \twelveCO/\CeighteenO\ ratios 
are observed in the $J$=2--1 transitions.
They may reflect subthermal excitation in the merger disk.

\end{enumerate}

\begin{figure}[h]
\epsscale{0.6}
 \plotone{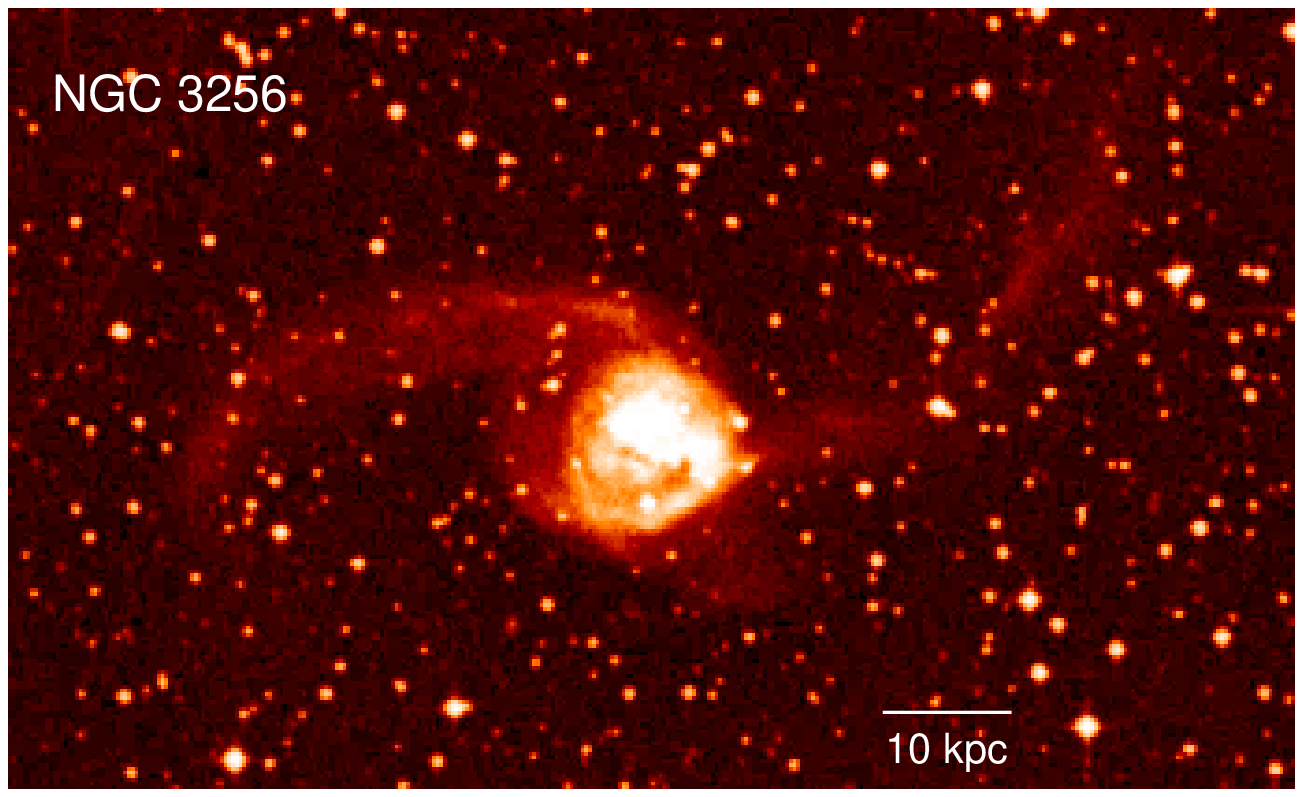}
\epsscale{0.365} 
 \plotone{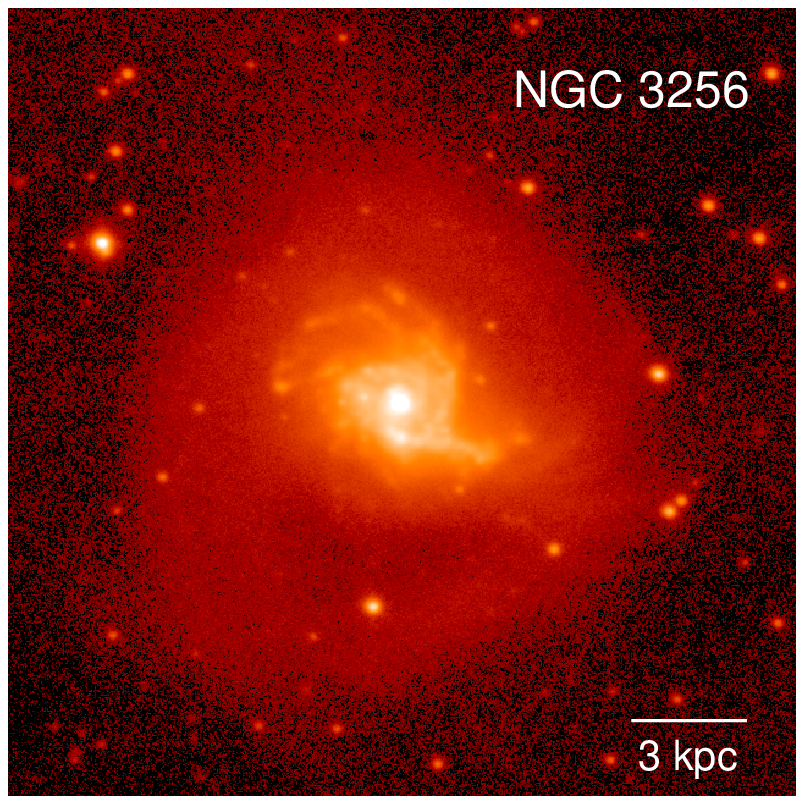} 
\epsscale{1.0}
\caption{Optical and near-IR images of NGC 3256.
(left) Blue-band image from the UK Schmidt Survey through the Digitized Sky Survey.
The image is 10\arcmin\ by 6\arcmin\ (= 100 kpc $\times$ 60 kpc).
(right) $K$-band image of NGC 3256 taken by \citet{Rothberg04}. The central 2\arcmin\ by 2\arcmin\ 
(= 20 kpc $\times$ 20 kpc) is shown. 
In each image, north is up and east is to the left, and intensity is shown with logarithmic scale.
    \notetoeditor{Please place these images horizontally side by side.}
\label{fig.opt} }
\end{figure}

\begin{figure}[h]
\epsscale{0.8}
\begin{center}
\plottwo{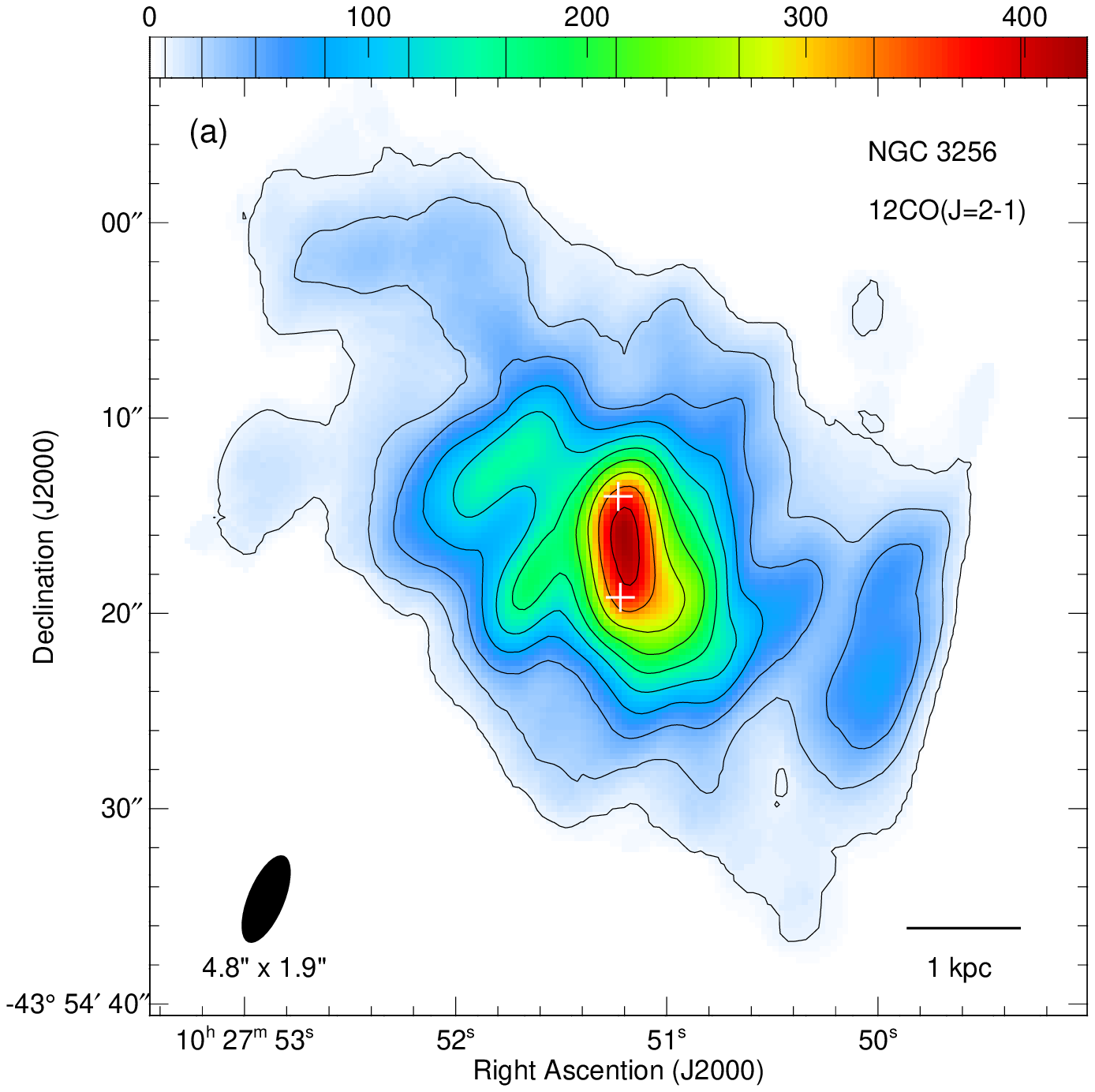}{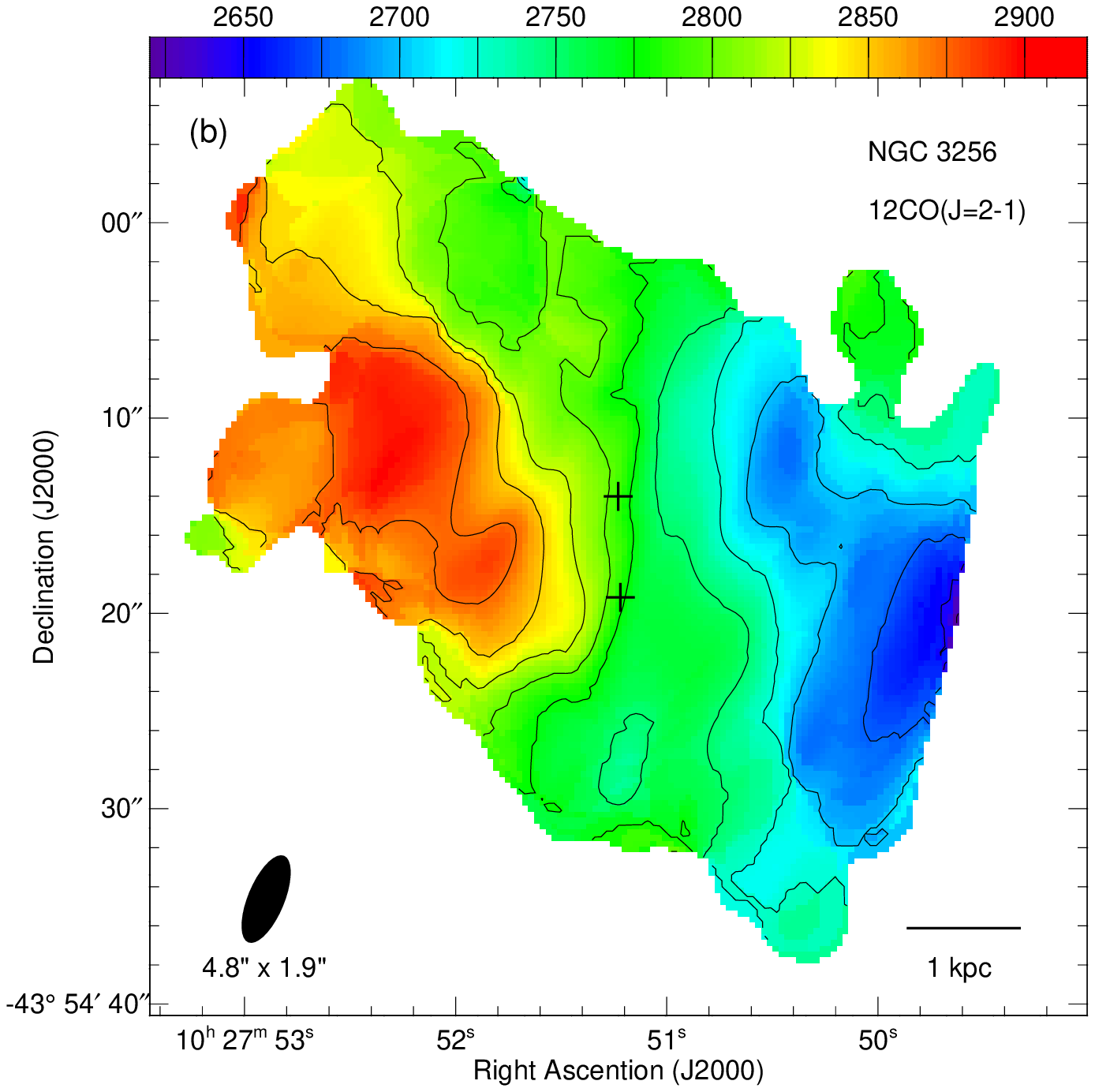} \\
\plottwo{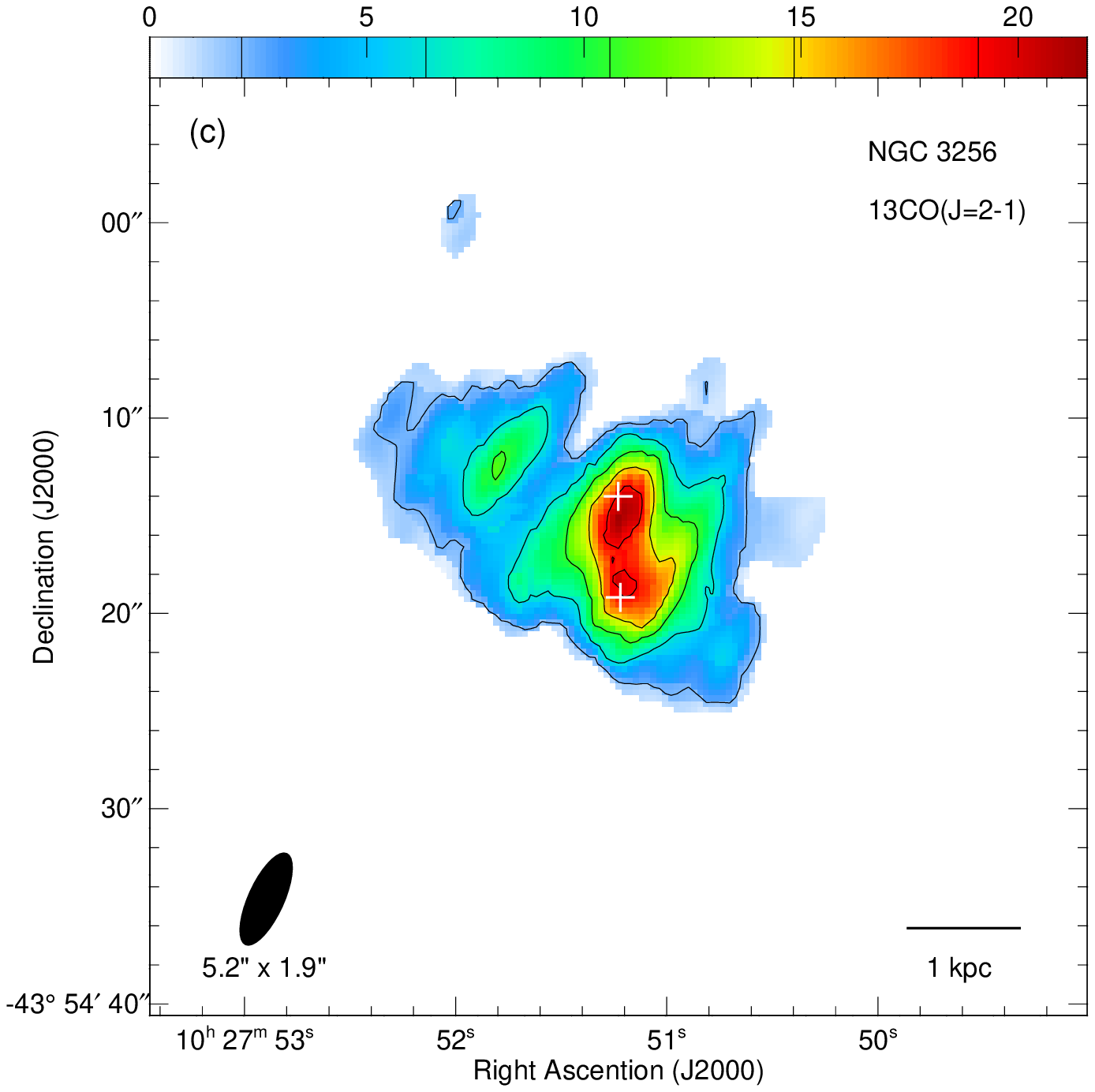}{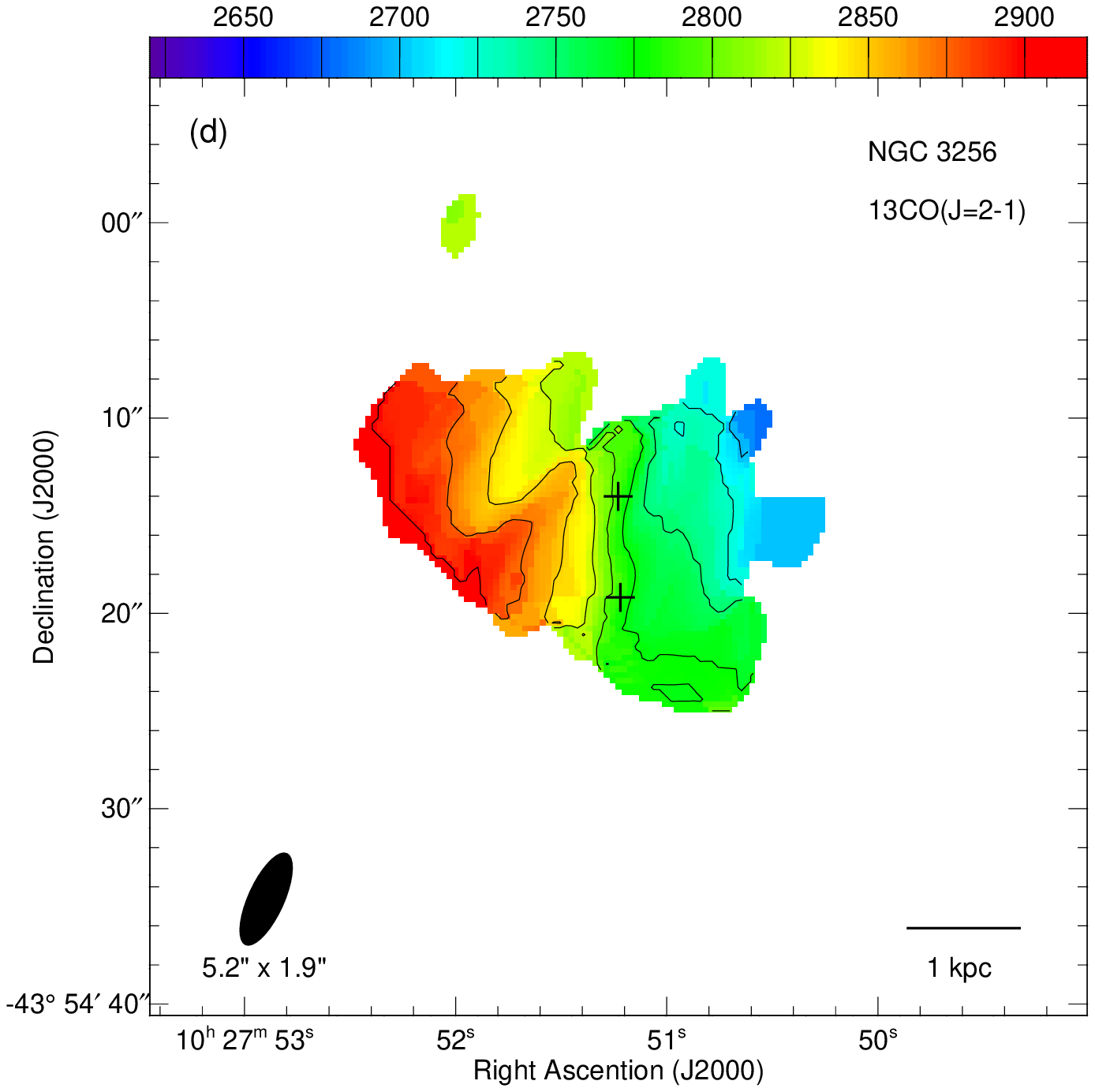} \\
\vspace{2mm}
\plottwo{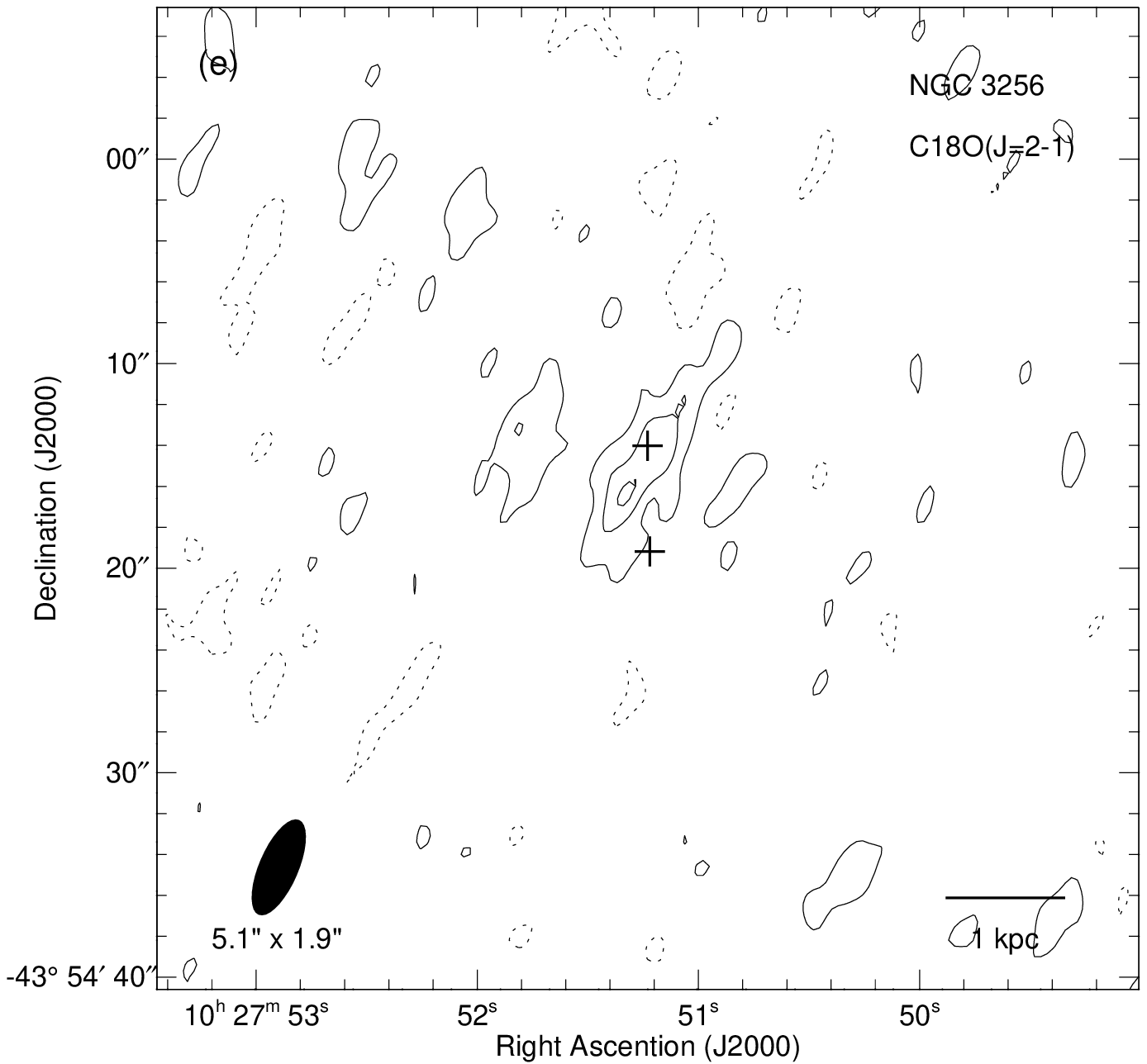}{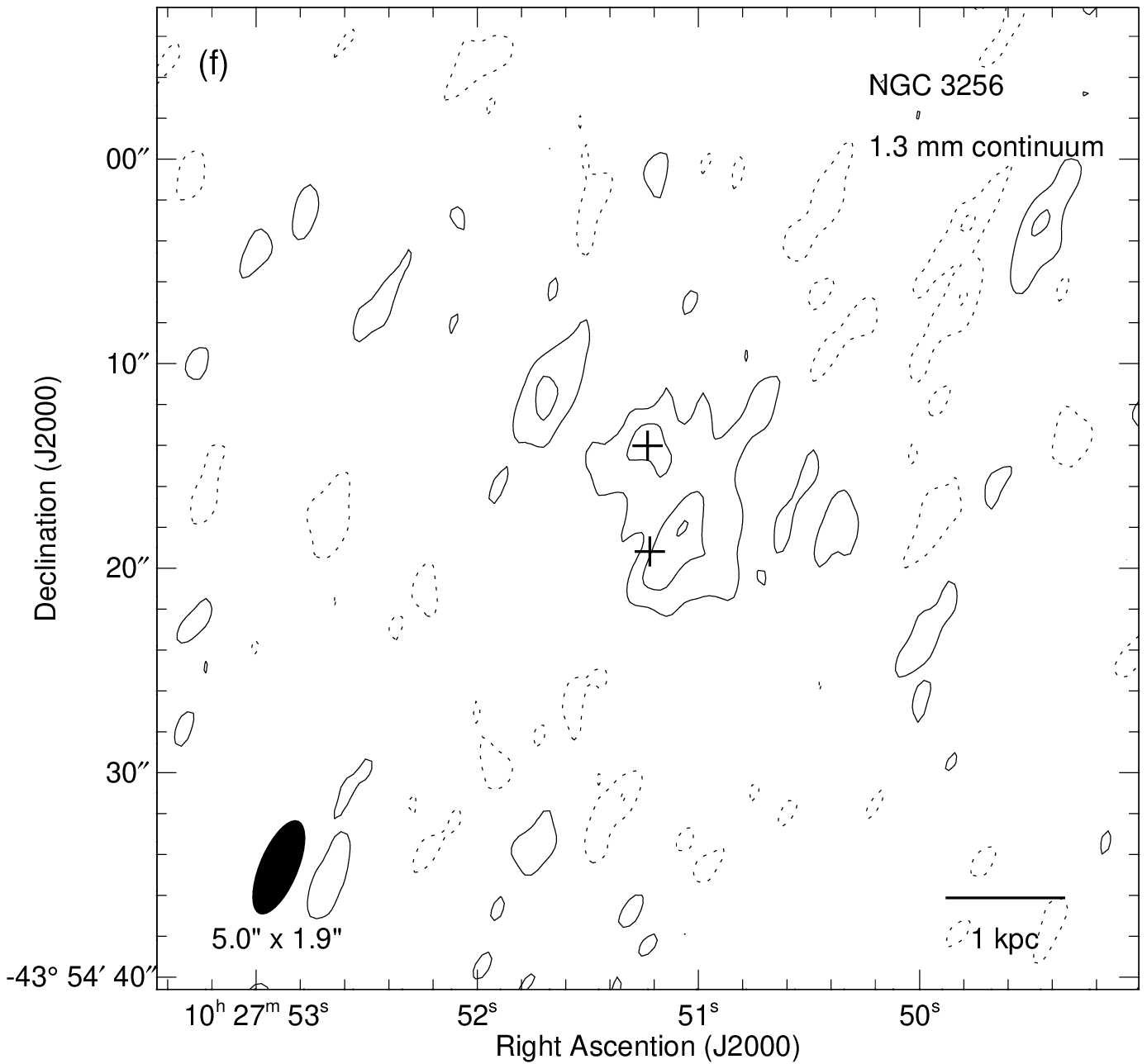} 
\end{center}
	\vspace{-5mm}
\caption{\small Natural-weighting maps of CO line and continuum emission in the center of NGC 3256.
(a) \twelveCO(2--1) integrated intensity. 
The $n$-th contour is at $7.1\times n^{1.75}$ Jy \perbeam\ \kms.
(b)  \twelveCO(2--1) mean velocity (i.e., 1st moment).
The velocity contours are in steps of 25 \kms.
(c) \thirteenCO(2--1) integrated intensity.
Contours are at $[1, 3, 5, 7, 9] \times 2.1$ Jy \perbeam\ \kms.
(d) \thirteenCO(2--1) mean velocity.
(e)  \CeighteenO(2--1) total intensity map, integrating from 2650 \kms\ to 2910 \kms.
Contours are in steps of 3.2 Jy \perbeam\ \kms\ ($= 2 \sigma$). 
(f) 1.3 mm (222.7 GHz) continuum map made by combining the data from both sidebands.
Contours are in steps of 5.0 mJy \perbeam\ ($= 2 \sigma$).
In each panel, the double nucleus is marked with a pair of crosses. 
The one in the north is the nucleus N and the other in the south is the nucleus S.
The synthesized beam of each map is shown at the bottom left corner.
\label{fig.naturalmap} }
\end{figure}

\begin{figure}[h]
\epsscale{0.5}
\plotone{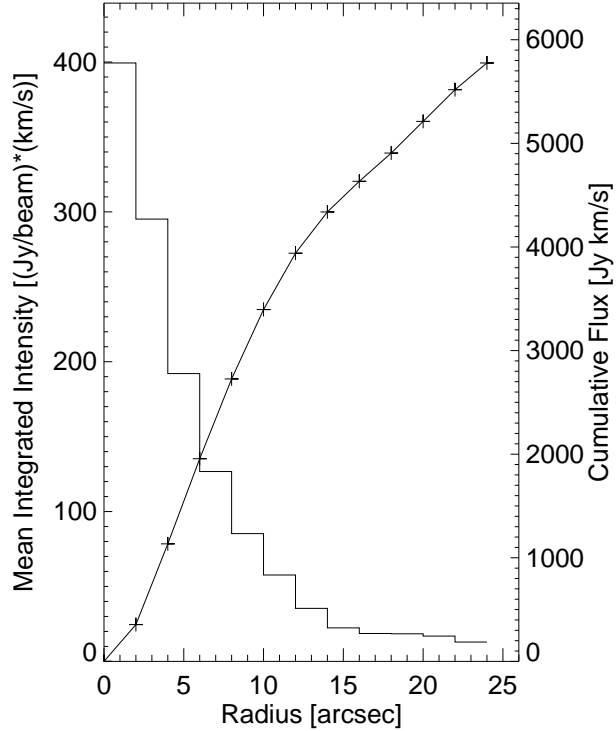} 
\epsscale{1.0}
\caption{Inclination-corrected radial distribution of \twelveCO(2--1) emission.
The mean integrated intensities (left axis) are measured in concentric rings in the galaxy plane,
for which we adopt an inclination of 45\arcdeg, a position angle of the major axis of 90\arcdeg,
and the center at the midpoint of the two nuclei. 
The geometry is consistent with the kinematical fit of our CO velocity data.
The cumulative flux (right axis) is calculated from the integrated intensities.
The line flux is measured from the integrated intensity map made by summing up
channel maps, rather than a moment map.
The natural weighting dataset is used for this plot after correcting for the primary beam attenuation.
The flux scale has a \tilde 10\% uncertainty.
\label{fig.12co.iring} }
\end{figure}

\begin{figure}[h]
\epsscale{1.0}
\plottwo{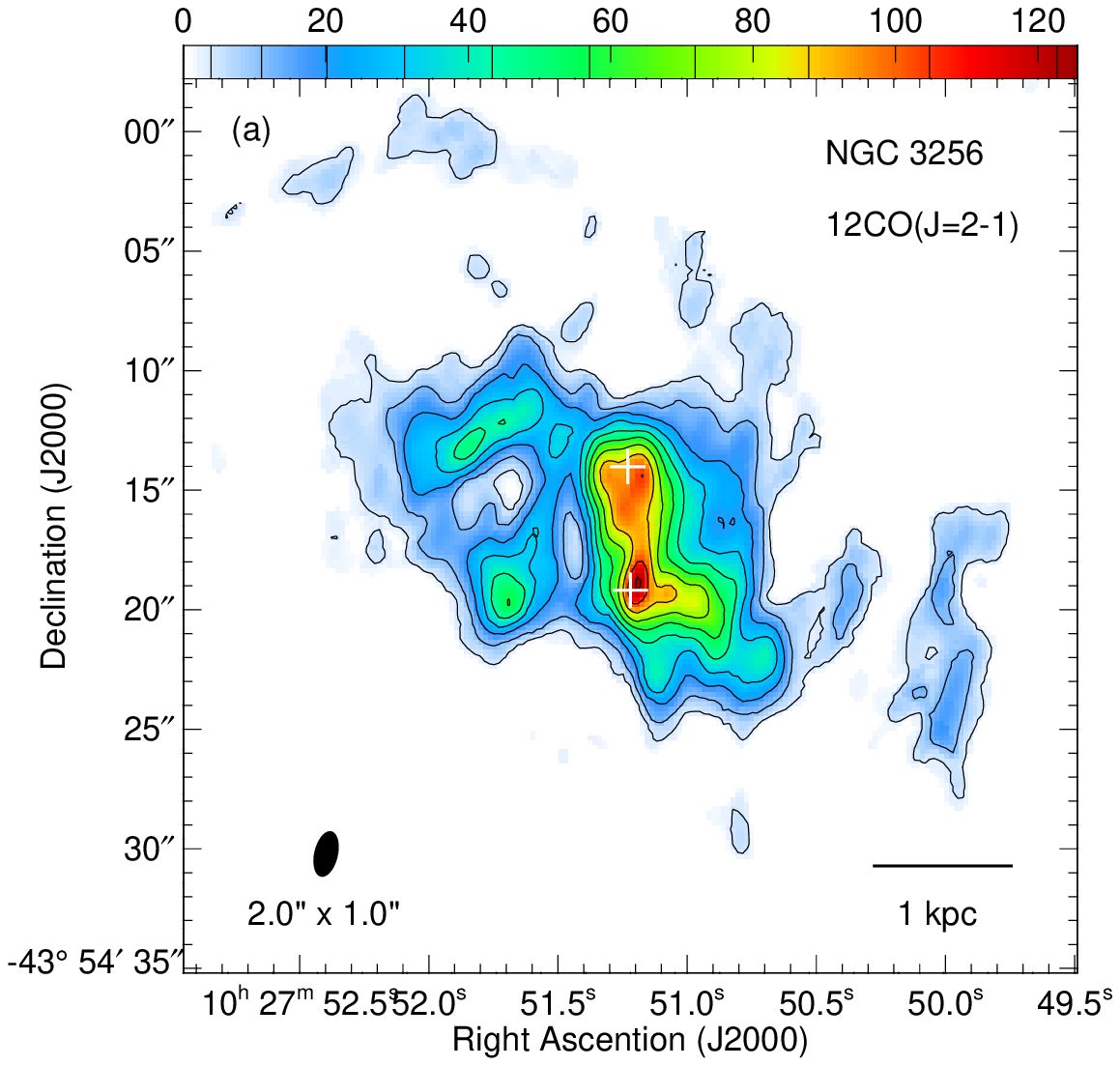}{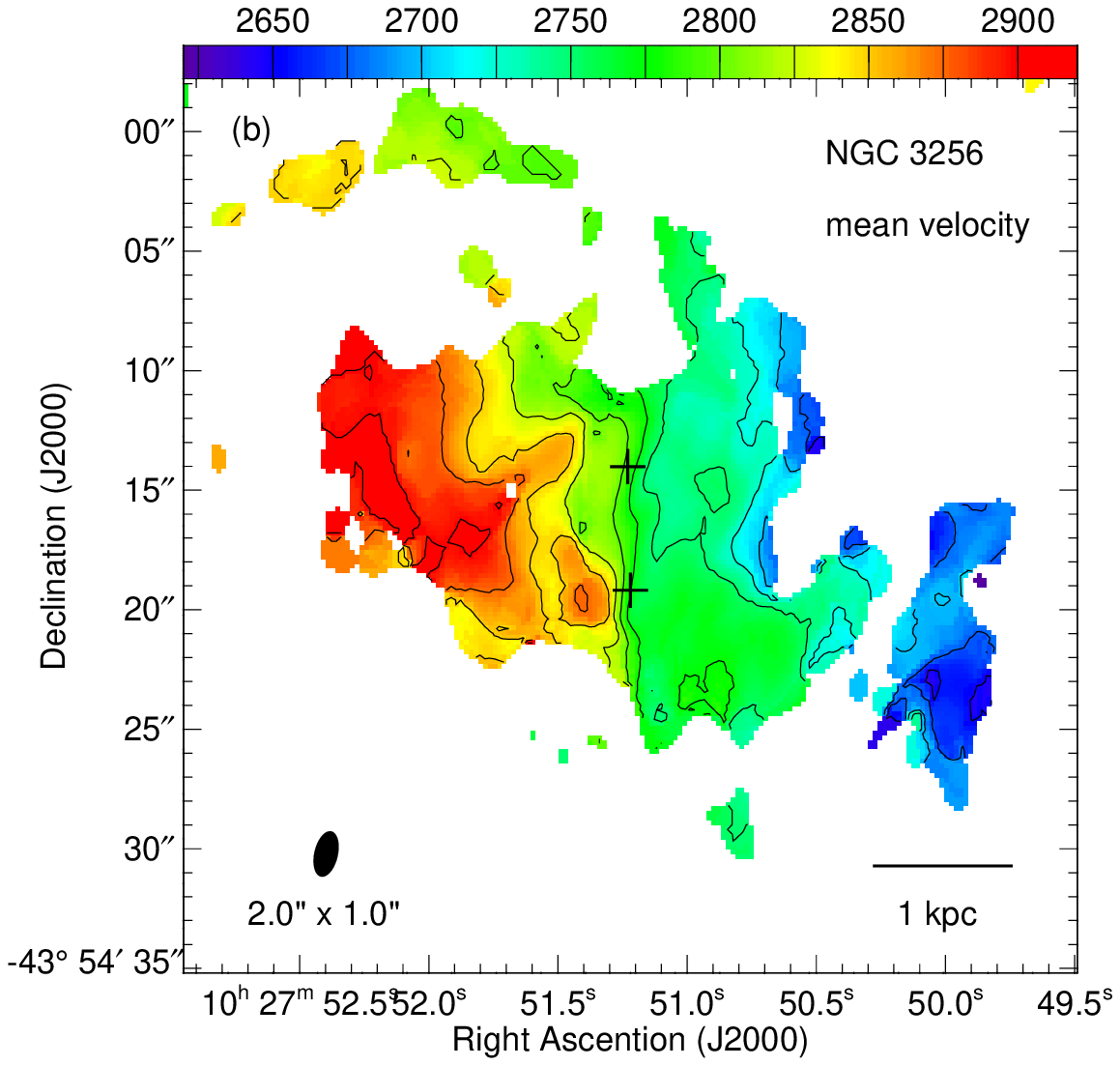} \\
\plottwo{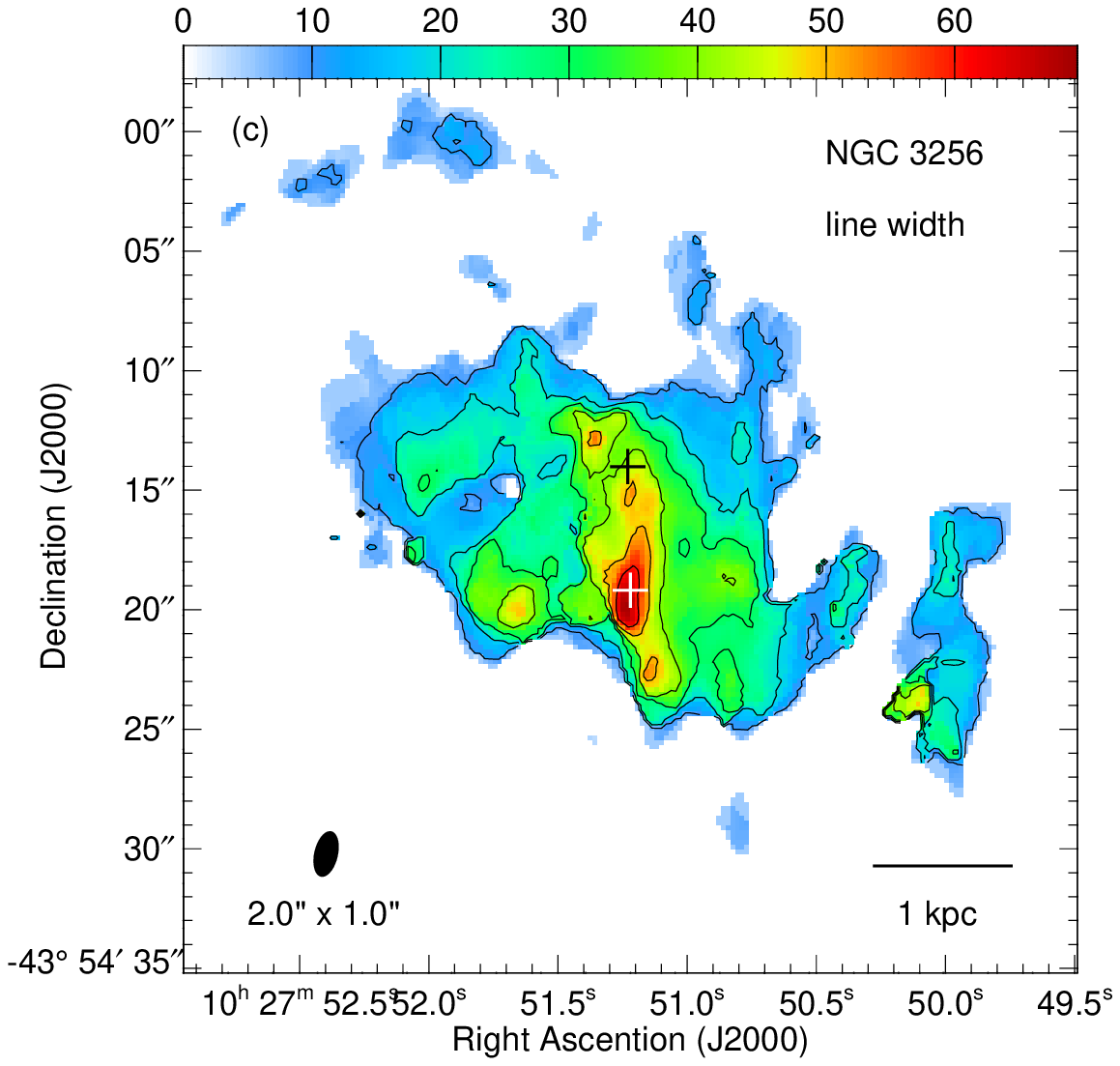}{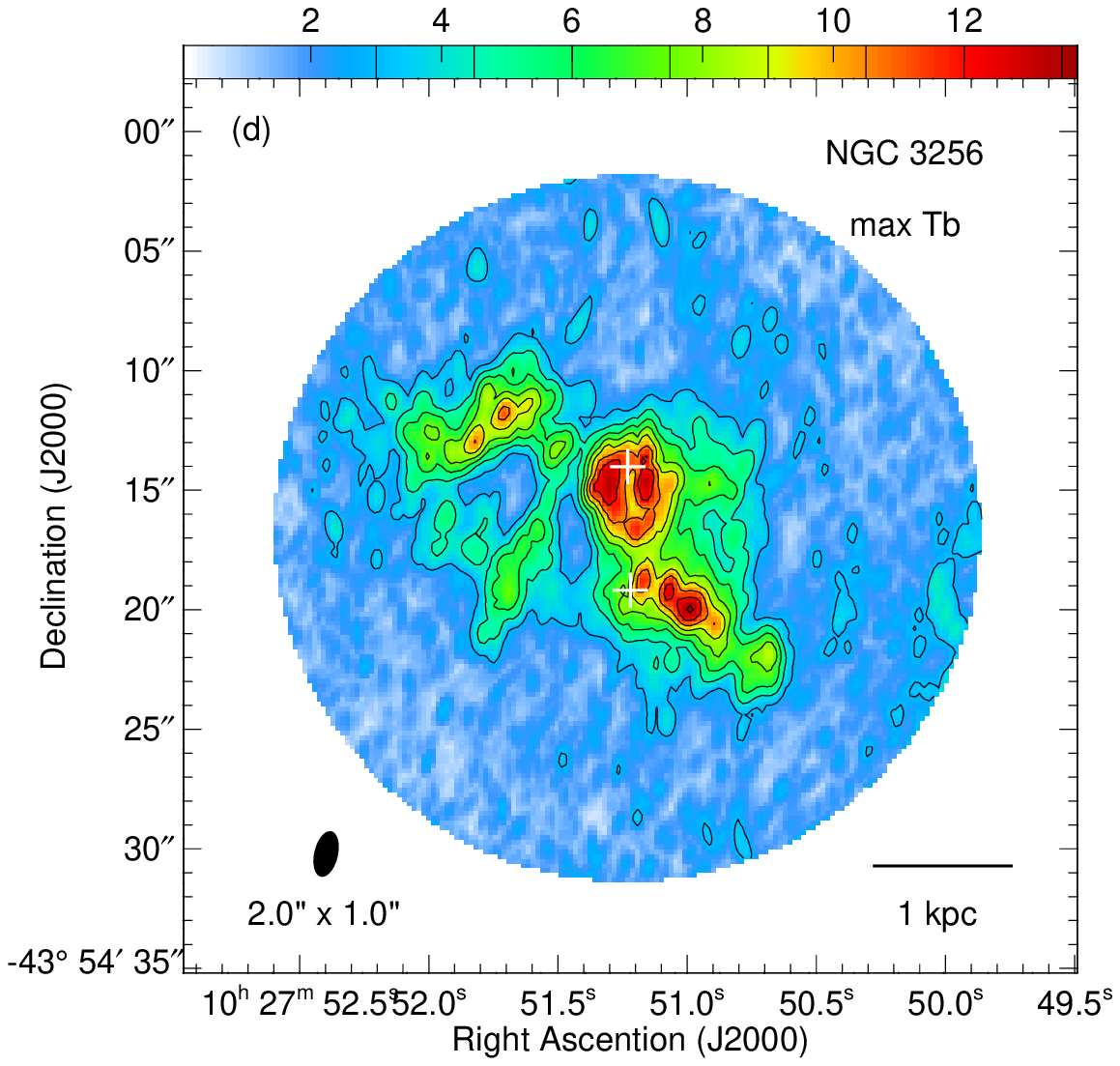} 
\caption{High-resolution maps of \twelveCO(2--1) emission in the center of NGC 3256. 
(a) Integrated intensity in unit of Jy  \perbeam\ \kms.
(b) Mean velocity in unit of \kms.
(c) Line width (i.e., the 2nd moment) in unit of \kms.
(d) Peak brightness temperature in unit of K.
The two nuclei of the merger are marked with $+$ signs.
The synthesized beam is at the bottom left corner.
The $n$-th contour in the integrated intensity map is at 
$3.9\times n^{1.5}$ Jy \perbeam\ \kms. 
The velocity contours are in steps of 25 \kms\ and 10 \kms\
in the 1st and 2nd moment maps, respectively.
The temperature contours are in steps of 1.5 K starting from 3 K.
The temperature map, only to which the primary beam correction is applied, is
masked in its noisier outer region.
The high-resolution data cube was made with super-uniform weighting 
and recovered 54 \% of the total CO flux in our natural weighting data shown in Fig. \ref{fig.naturalmap}.
\label{fig.cor-2s} }
\end{figure}

\begin{figure}[h]
\epsscale{1.0}
\plotone{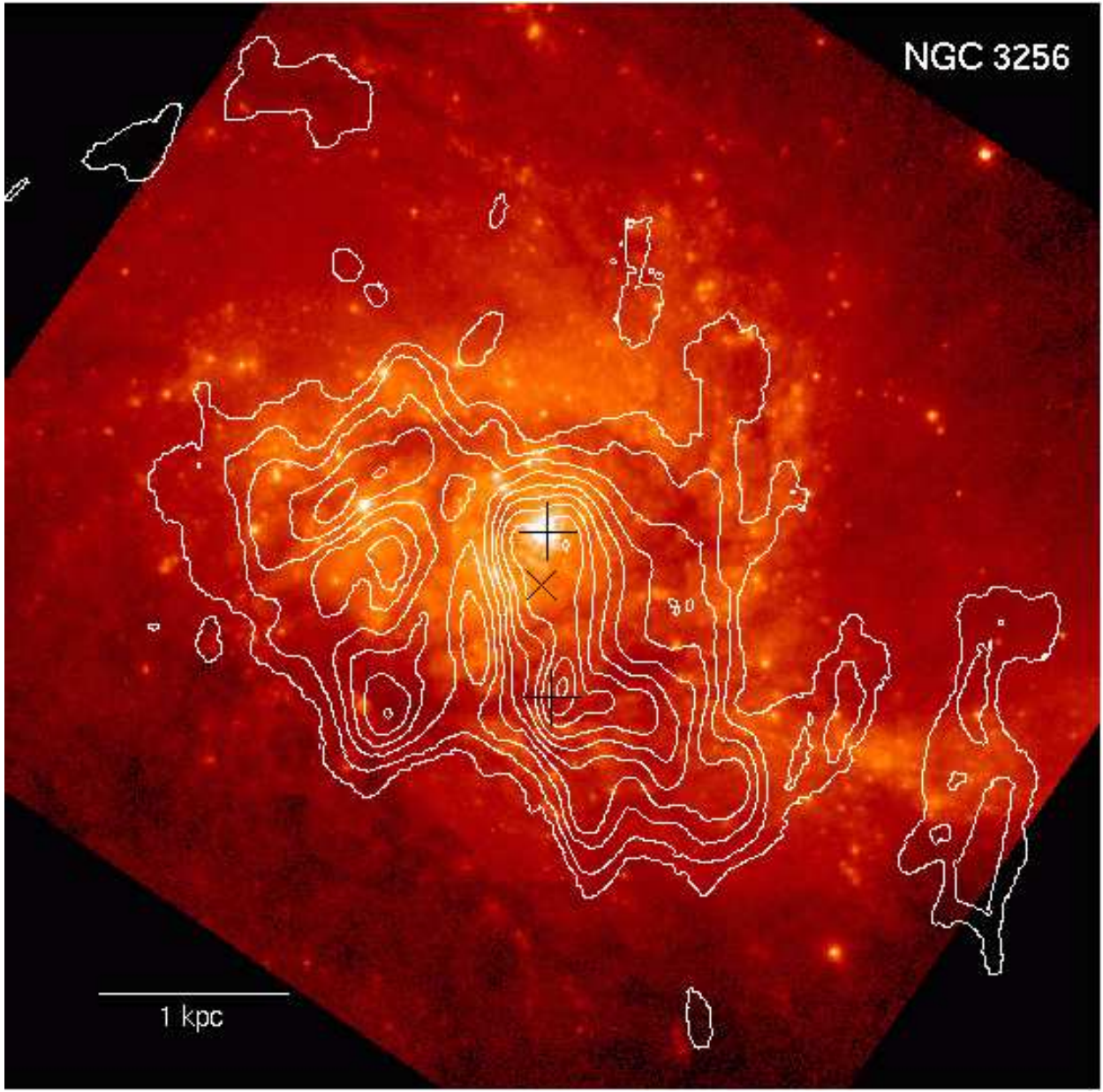}
\caption{SMA CO(2--1) contours overlaid on an HST {\it I}-band image of the
center of NGC 3256.
The CO contours are the same as in Fig. \ref{fig.cor-2s}, and the
optical image is shown with logarithmic scale.
The radio positions of the two nuclei are marked with  plus signs.
The $\times$ between the two nuclei is the dynamical center of the galaxy 
estimated from the CO velocity field.
North is up, and east is to the left. 
This figure covers an area of 34\arcsec\ by 34\arcsec.
See \citet{Zepf99} for the HST observations and a color composite image of the same area.
\label{fig.coonhst} }
\end{figure}

\begin{figure}[h]
\epsscale{1.0}
\plottwo{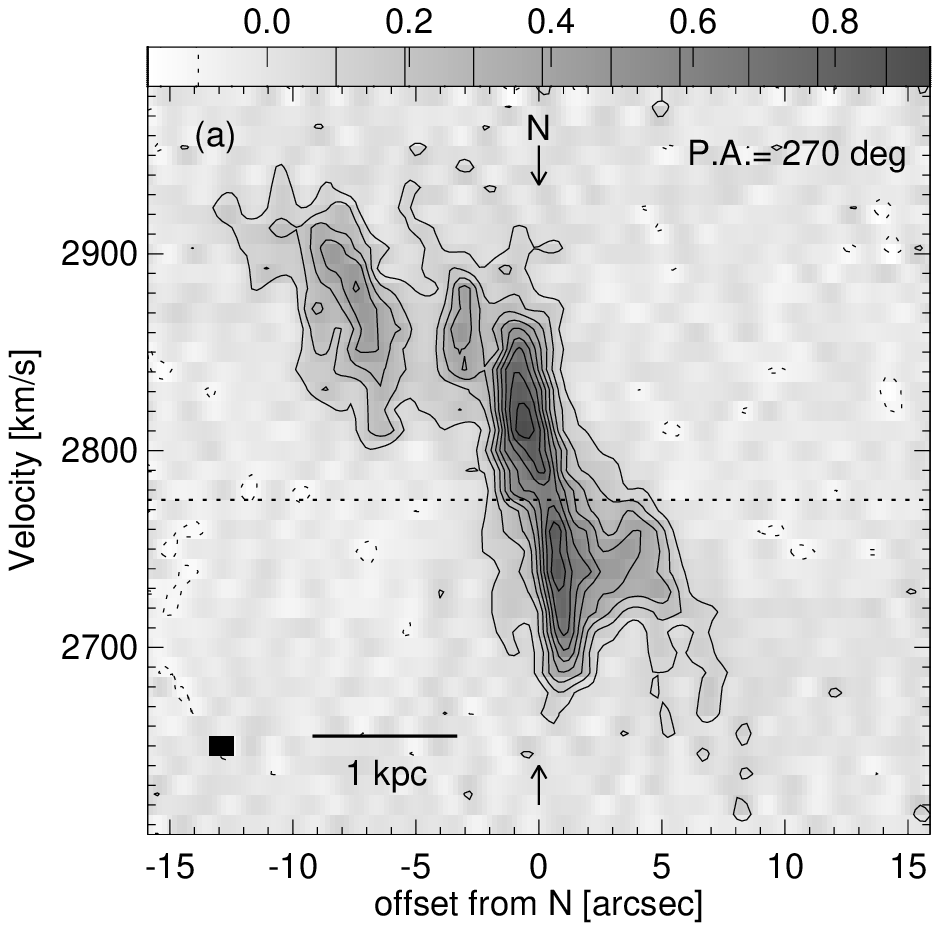}{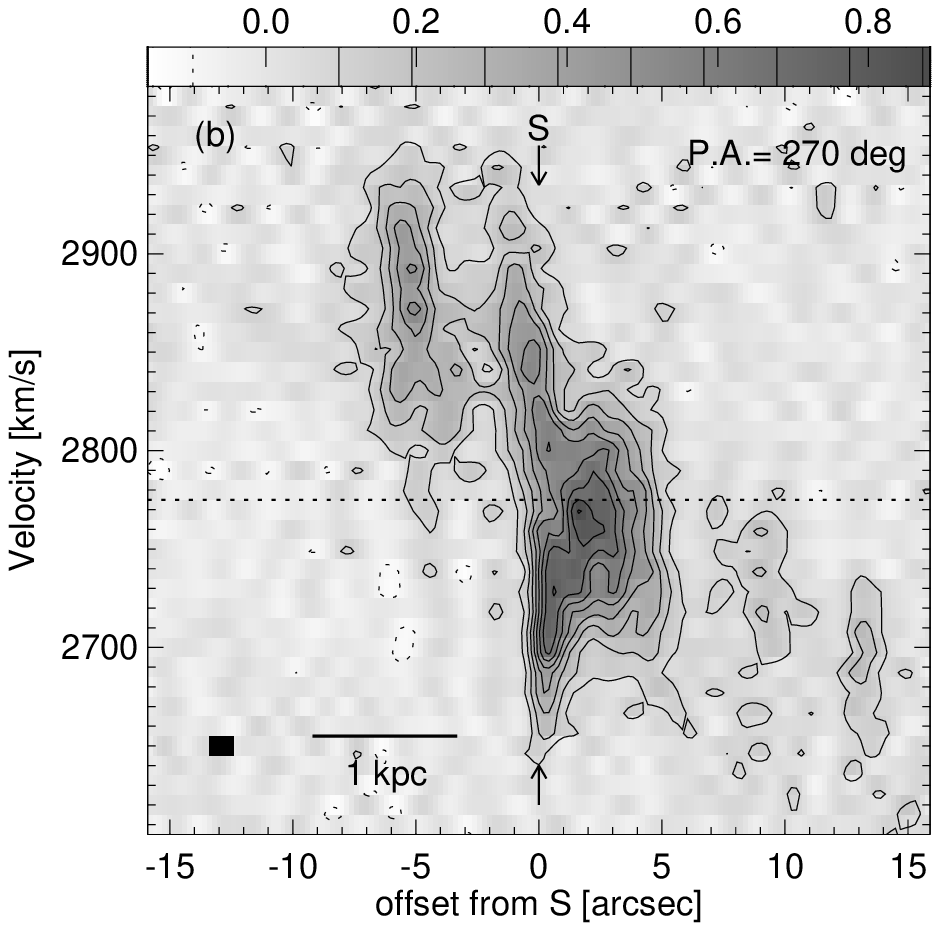}  \\
\plottwo{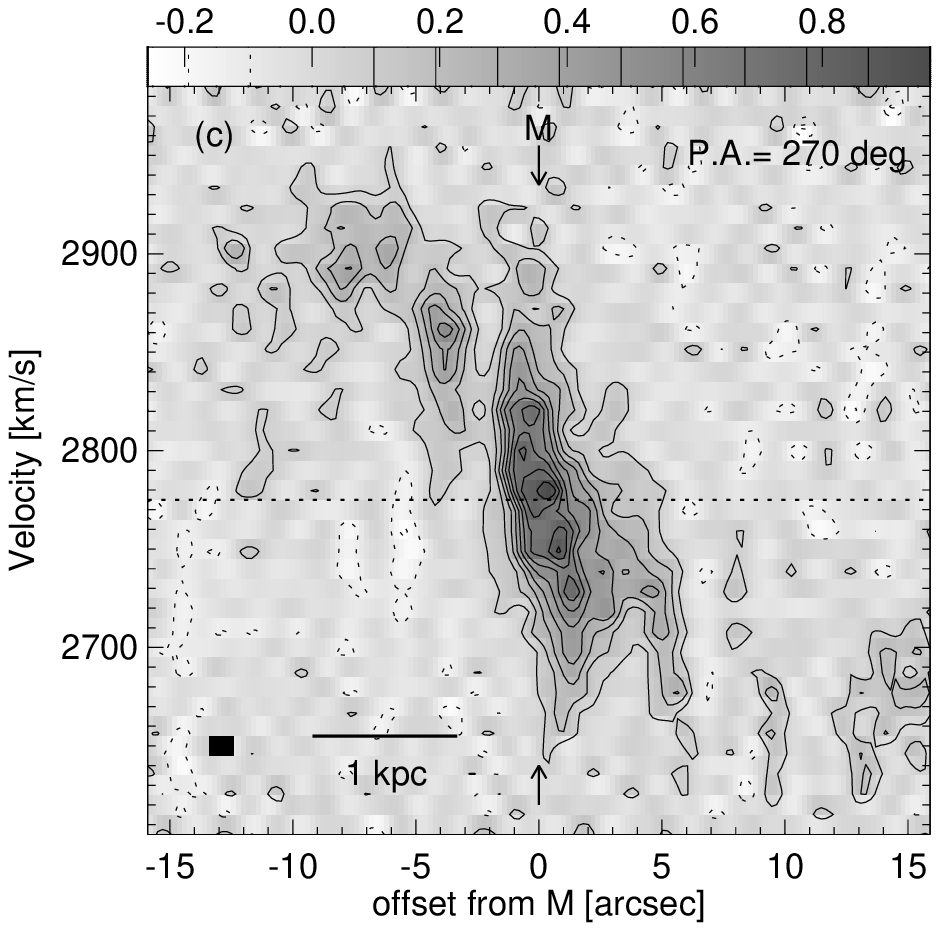}{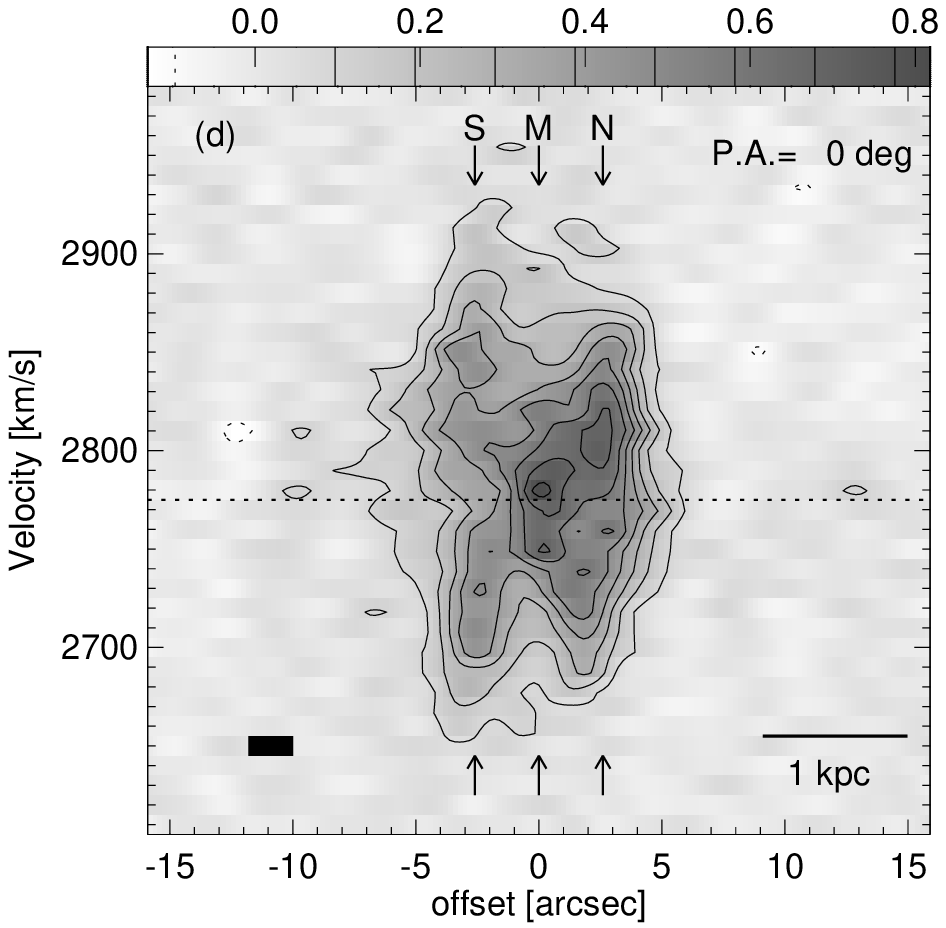}
\caption{Position-velocity (PV) diagrams made from the high-resolution data cube.
(a) PV diagram along P.A.$=  270$\degr\ across the N (north) nucleus. 
(b) PV diagram along P.A.$=  270$\degr\ across the S (south) nucleus.
(c) PV diagram along P.A.$=  270$\degr\ across the middle point (M) of the double nucleus.
(d) PV diagram along P.A.$=  0$\degr\  across the two nuclei.
Each PV cut has a 2\arcsec\ width except the one across M that has a 0\farcs2 width to minimize
the contamination by the gas around the N and S nuclei.
The black rectangle at the bottom-left corner of each panel shows the spatial resolution
along the PV cut and the 10 \kms\ velocity resolution of the data.
Contours are in steps of 97 mJy \perbeam\ or 1.2 K in the brightness temperature.
The dotted line in each diagram shows the systemic velocity of the galaxy ($V_{\rm sys} = 2775$ \kms).
\label{fig.high-res-pv} }
\end{figure}

\begin{figure}[h]
\epsscale{0.7}
\plotone{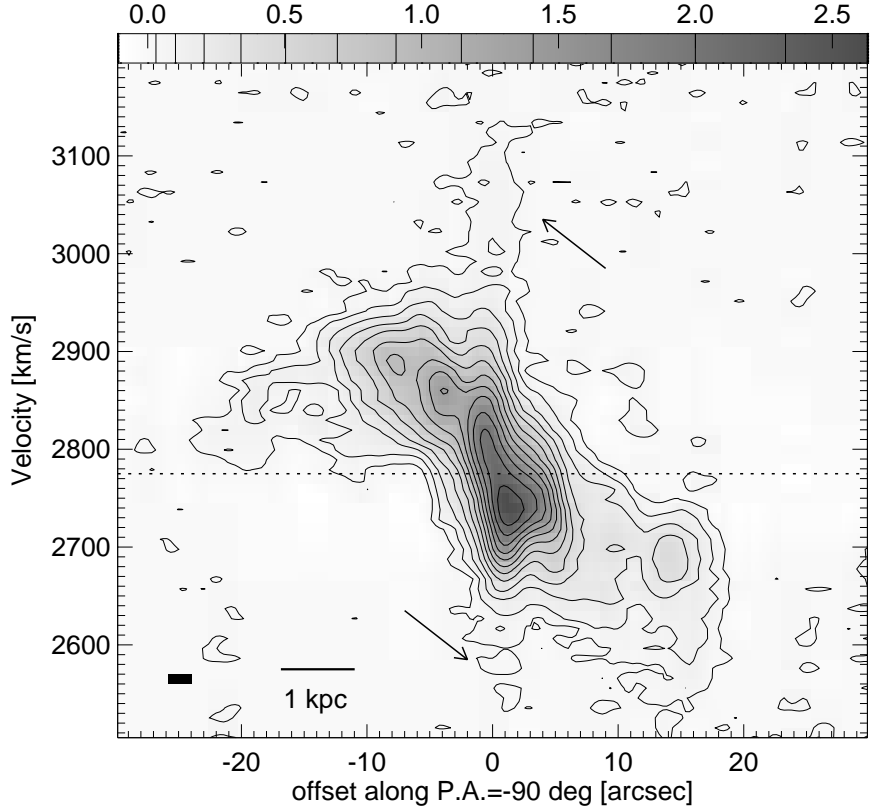}
\caption{Position-velocity diagram along P.A.$= -90$\degr.
The dotted line indicates the systemic velocity of the galaxy, for which we adopt 
$V_{\rm sys}$(radio, LSR) $= 2775$ \kms.
The zero point of the horizontal axis is where the double nucleus is.
Intensity is in Jy \perbeam\ and can be converted to brightness temperature in kelvin by multiplying 2.8.
The $n$-th contour is at $n^{1.75} \times 30$ mJy \perbeam.
The black rectangle at the bottom left corner tells the spatial and velocity resolutions.
The arrows point the high velocity gas in the galactic center.
\label{fig.majpv} }
\end{figure}

\begin{figure}[h]
\epsscale{1.0}
\plottwo{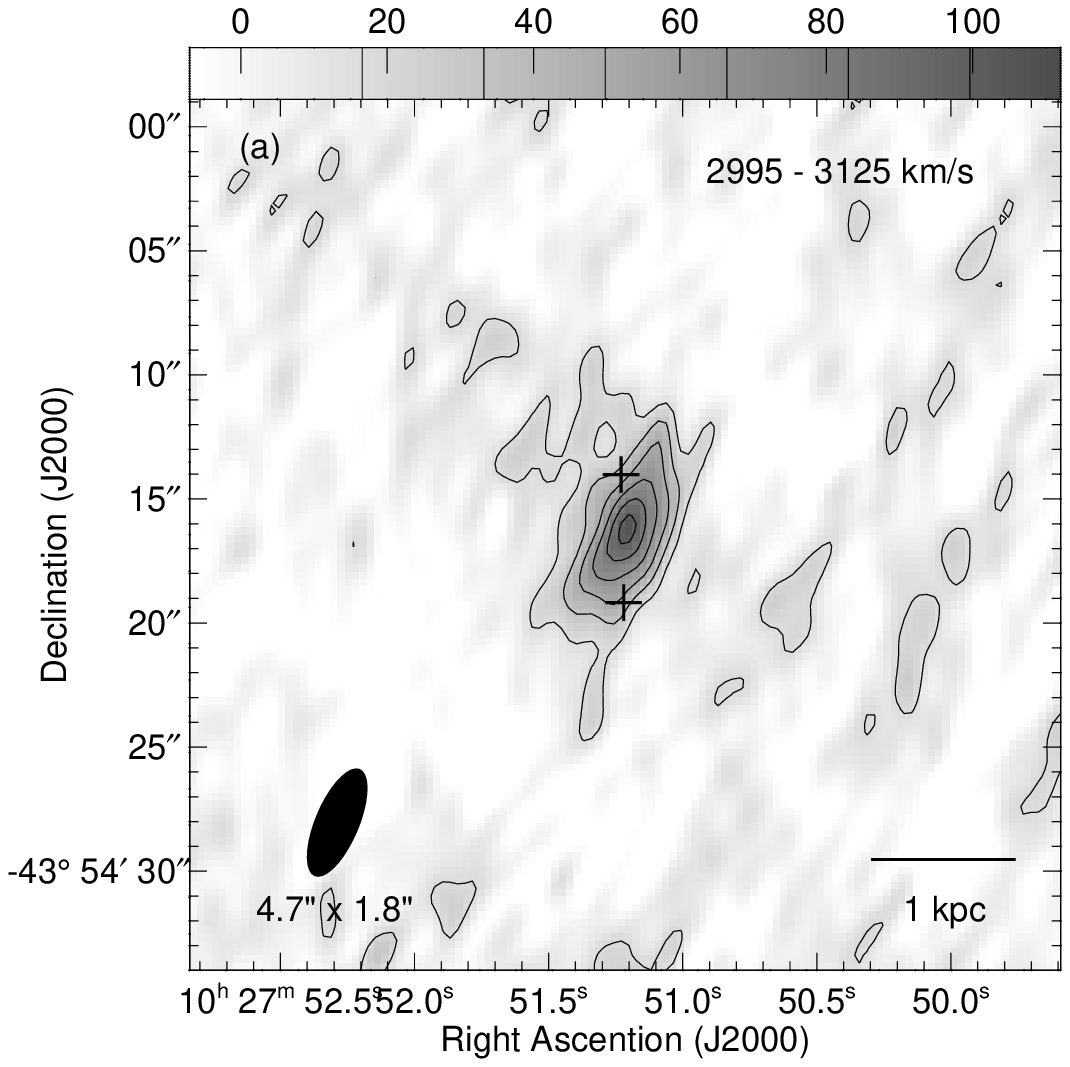} {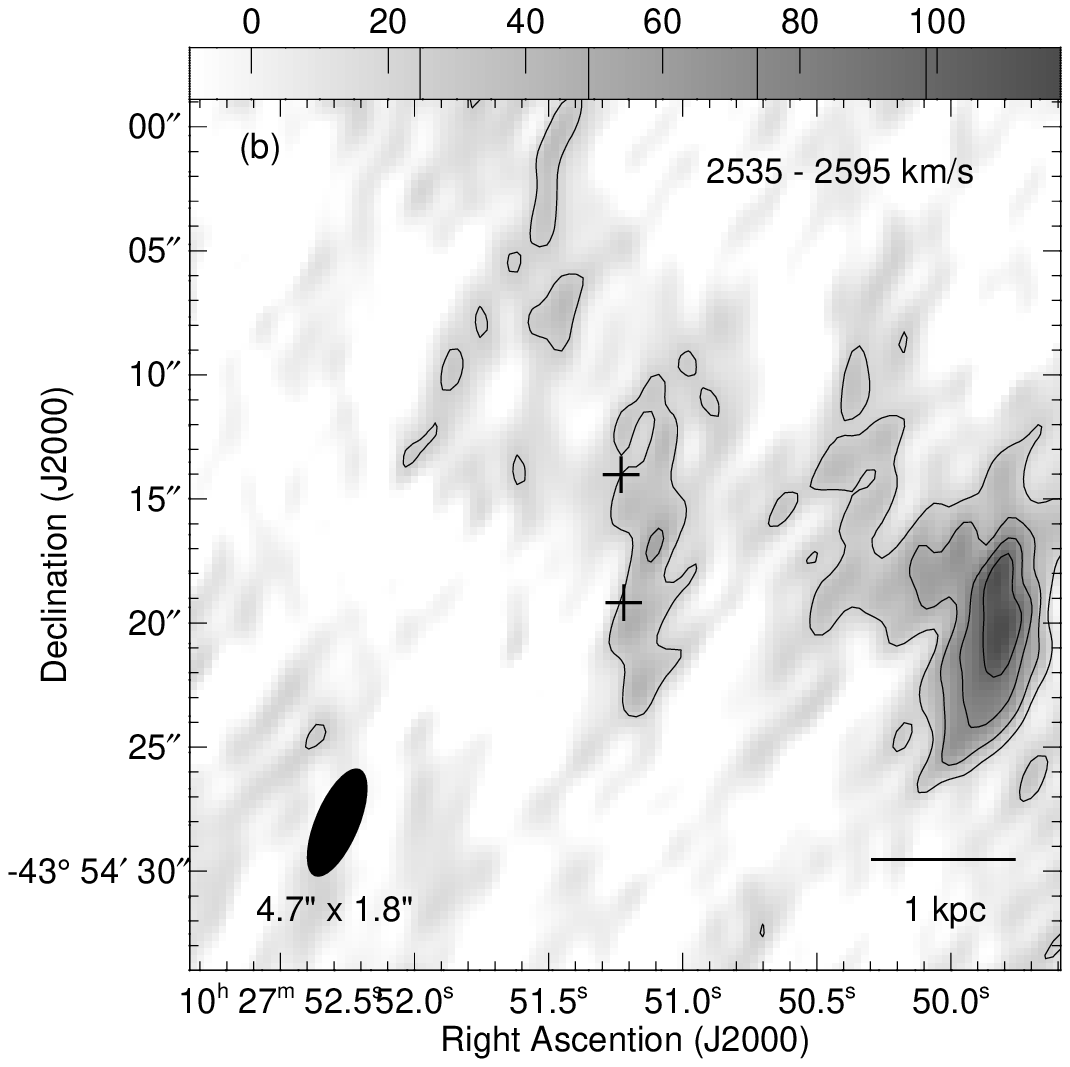}
\caption{Molecular gas at extreme velocities.
The maps average velocities where either redshifted ($a$) or blueshifted ($b$) high velocity
gas is seen in the galactic center.
Contours are in steps of 2 $\sigma$ in each map, i.e., 
16.6 mJy \perbeam\ for the redshifted map and 24.6 mJy \perbeam\ for the blueshifted map.
The plus signs mark the two radio nuclei.
The synthesized beam is on the bottom left corner of each panel.
The arc-line emission 15\arcsec\ west of the nucleus in the blueshifted map is
not the high velocity gas but corresponds to a spiral feature in the outer disk, where noncircular 
motion is significant.
\label{fig.blue-red-gas} }
\end{figure}

\begin{figure}[h]
\epsscale{0.7}
\plotone{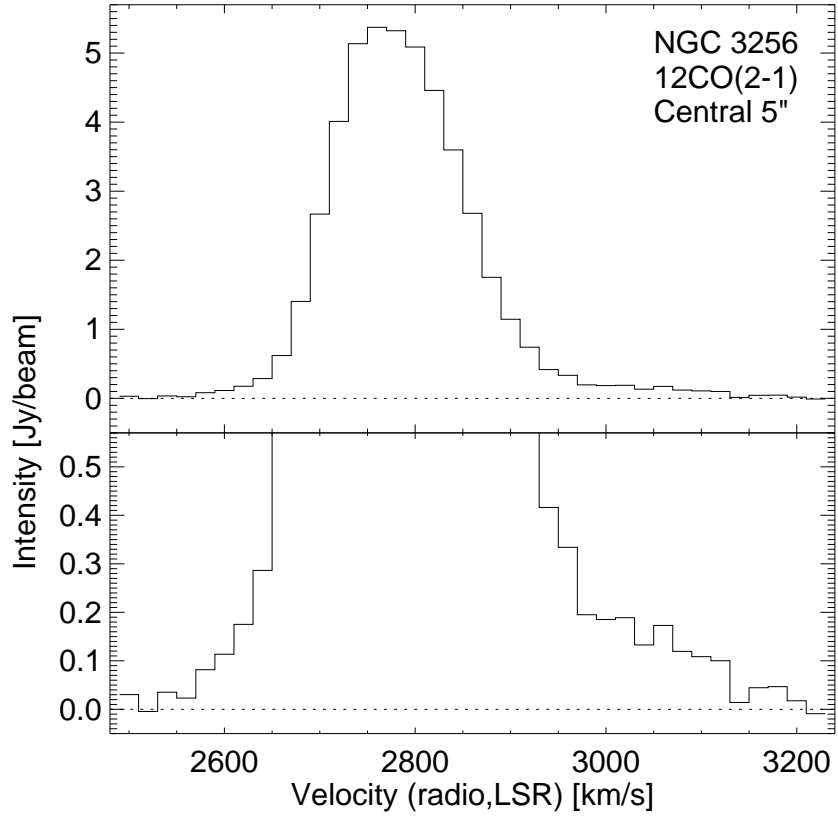}
\epsscale{1.0}
\caption{Spectrum of \twelveCO(2--1) emission in the central 5\arcsec\ of NGC 3256.
The bottom frame is to show the wing emission from the high velocity gas.
The spectrum is the one at the midpoint of the binary nuclei.
The r.m.s. noise in this profile is 30 mJy \perbeam; 
the noise was measured in the image domain in the data cube from which the spectrum was obtained. 
The spectrum is from the natural-weighing data cube convolved to a 5\arcsec\ resolution (FWHM).
\label{fig.hvgspec} }
\end{figure}

\begin{figure}[h]
\epsscale{0.43}
\plotone{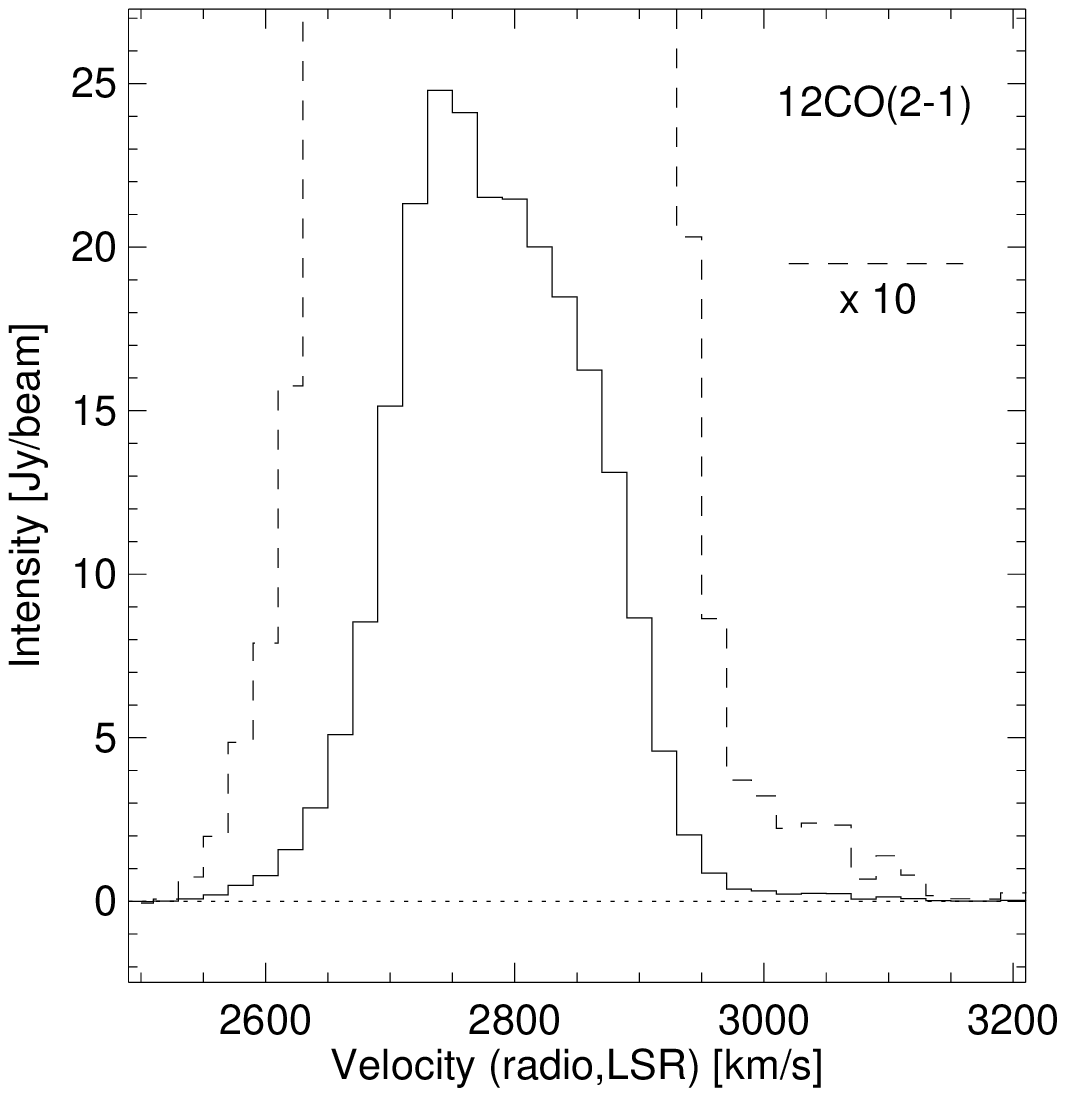} \\ \vspace{-4mm}
\plotone{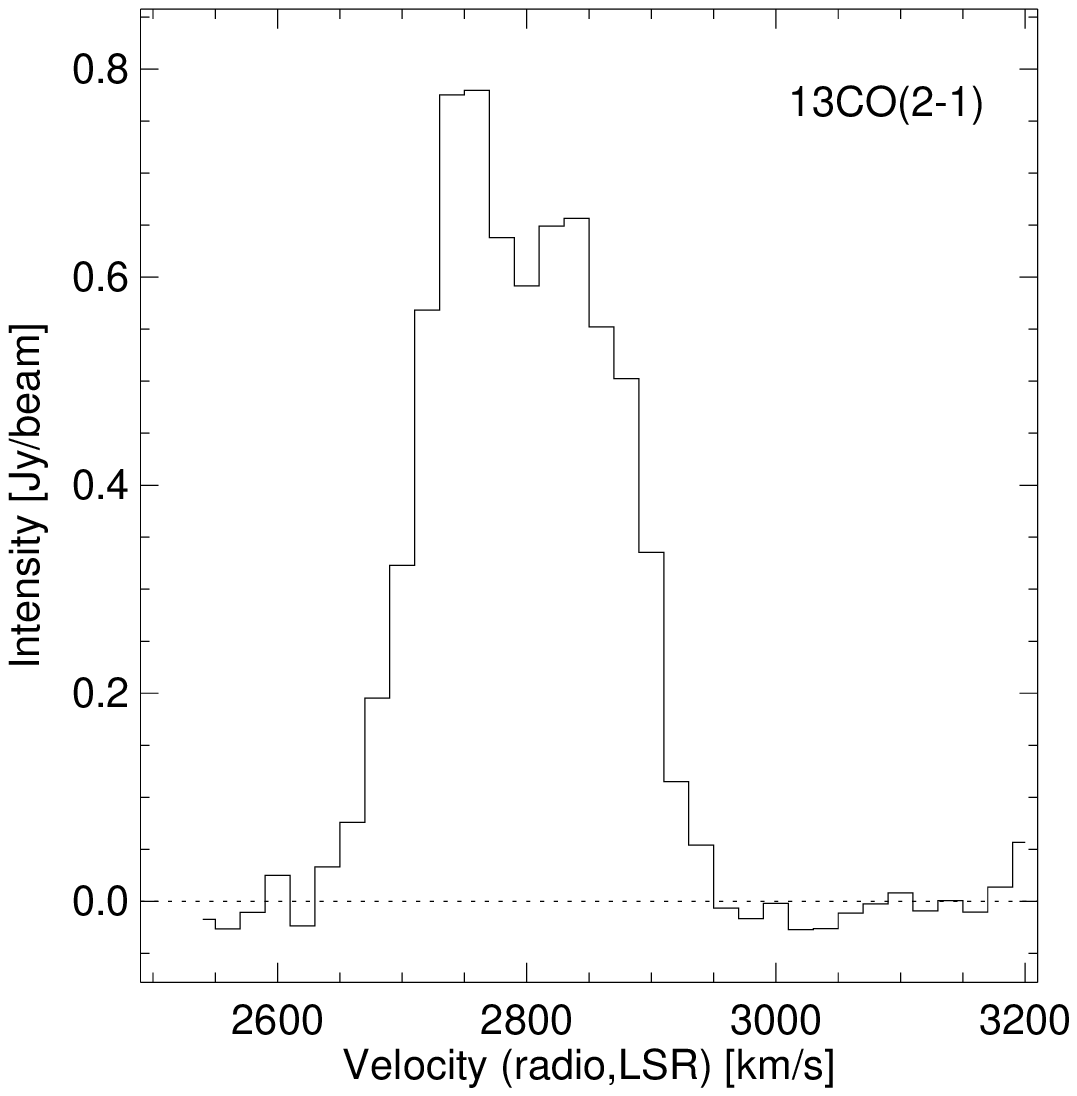} \\ \vspace{-4mm}
\plotone{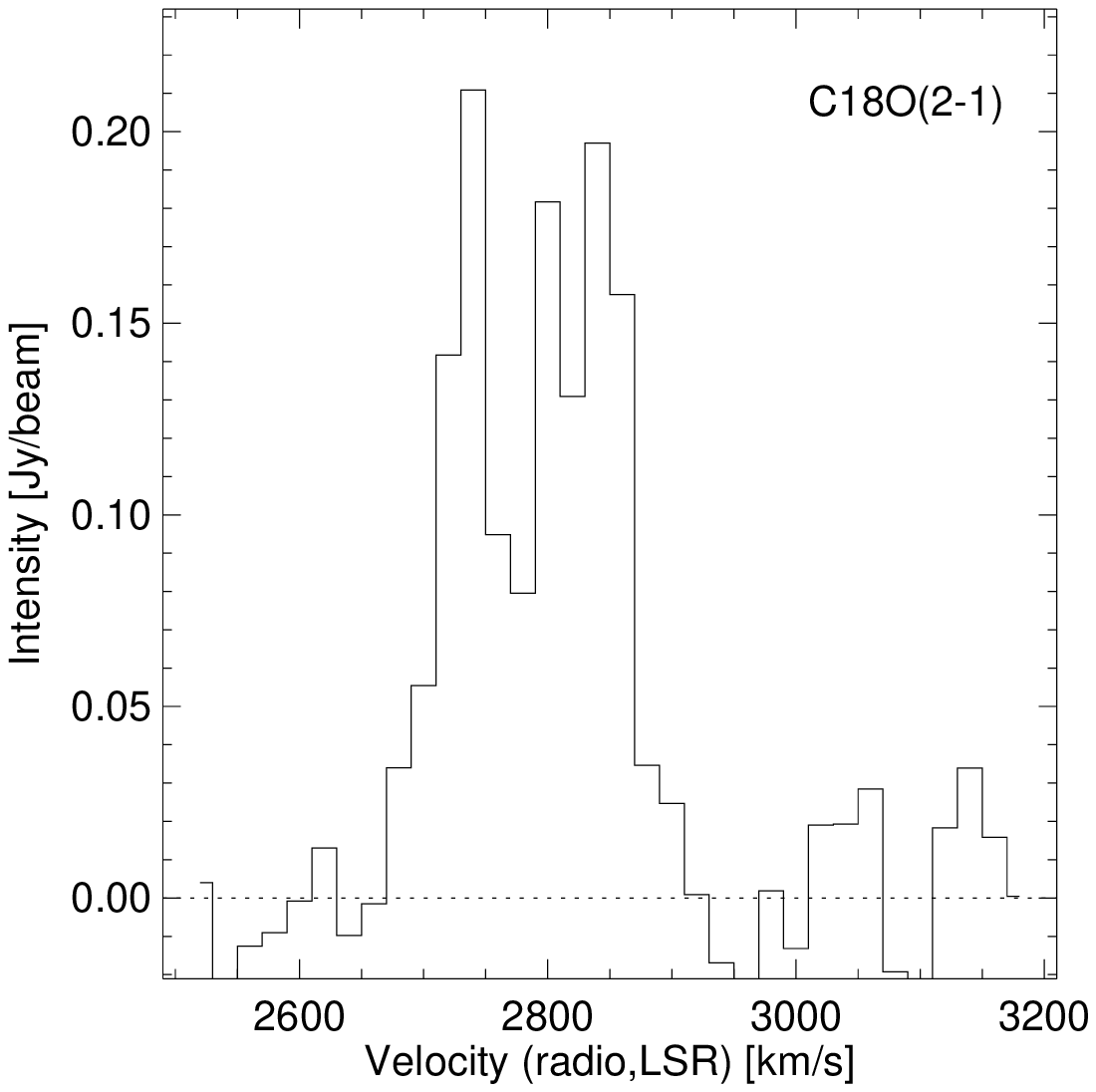} 
\epsscale{1.0}
\vspace{-2mm}
\caption{Spectra of the three CO lines in the central 30\arcsec\ of NGC 3256.
The natural-weighting data cube of each line was corrected for primary-beam attenuation and
then convolved to the 30\arcsec\ resolution (FHWM). 
The spectrum was then measured at the position of the phase tracking center, which is the middle
point of the double nucleus.
The dashed line in the \twelveCO\ panel is the same \twelveCO\ spectrum multiplied by 10. 
\label{fig.cospectra} }
\end{figure}

\begin{figure}[h]
\epsscale{0.6}
\plotone{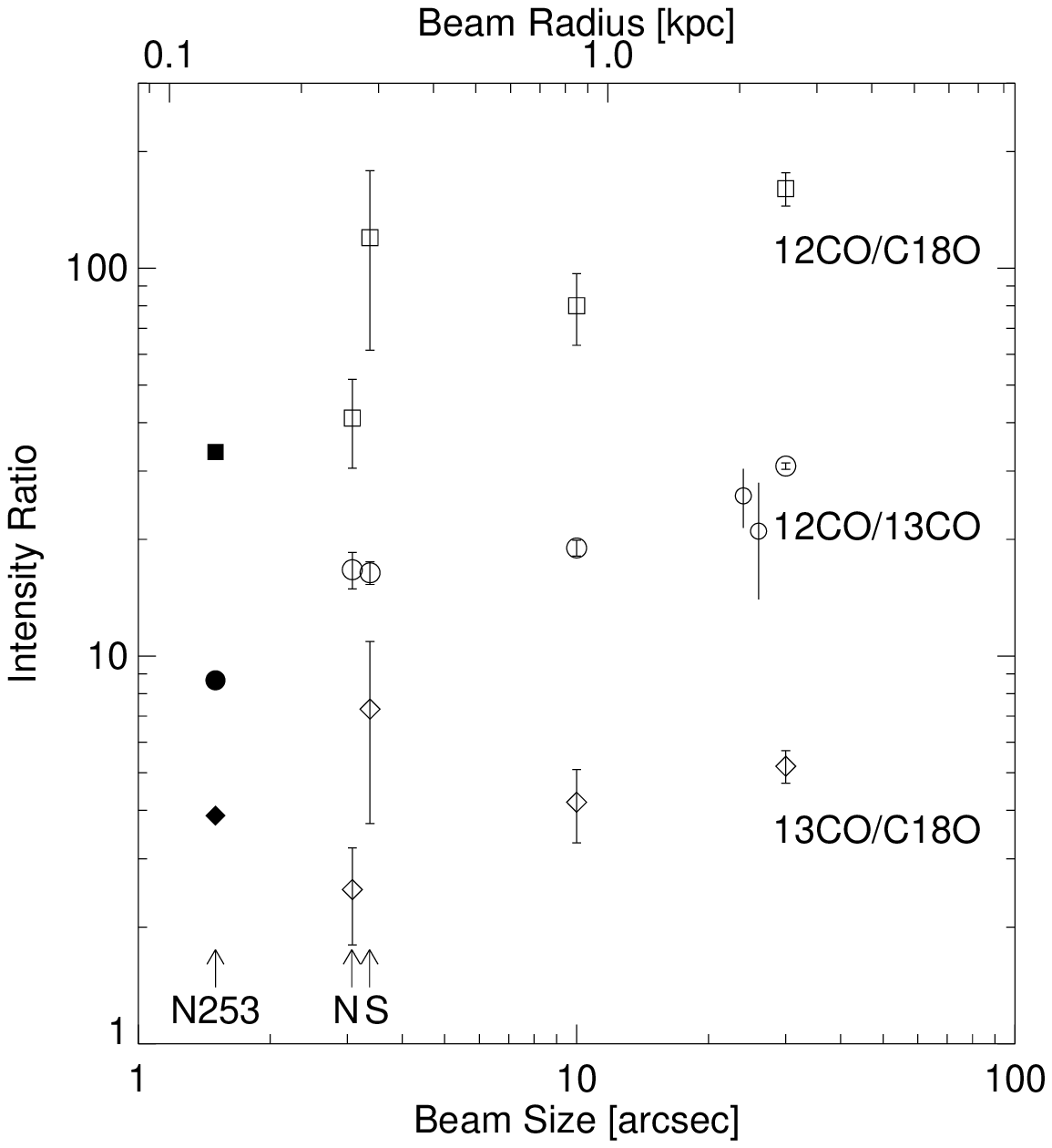} 
\epsscale{1.0}
\caption{Intensity ratio of the $J$=2--1 transitions of CO isotopomers in the centers of 
NGC 3256 (open symbols) and NGC 253 (filled symbols).
Circles, squares, and diamonds are for 
\twelveCO/\thirteenCO, \twelveCO/\CeighteenO, and \thirteenCO/\CeighteenO,
respectively.
The ratios in NGC 3256 are computed between brightness temperatures integrated over the range of 
2590 \kms\ -- 2970 \kms. 
The rightmost and middle data of NGC 3256 are measured at the midpoint of the double nuclei
from the natural weighting data convolved to 30\arcsec\ and 10\arcsec\ resolutions (FWHM), respectively.
The left group of NGC 3256 data are measured at the N nucleus (left points) and S nucleus (right points)
in the natural weighting data convolved to 5\farcs2 $\times$ 2\farcs0 resolution.
The nuclear data are assigned the beam size of 3.2 (=$\sqrt{5.2\times2.0}$) arcsec
and are given a slight horizontal offset for each nucleus to avoid overlap.
Correction for the primary beam attenuation was applied before each convolution.
Error bars are for $\pm 1 \sigma$.
The two \twelveCO/\thirteenCO\ data points for the beam size of about 25\arcsec\ are
from the single dish observations of \citet{Casoli92b} and \citet{Garay93}.
The data for the center of NGC 253 are from the SMA observations of \citet{Sakamoto06} convolved 
to a 250 pc = 15\arcsec\ (FWHM) resolution and integrated over the full velocity range of 50 -- 450 \kms. 
(The beam size of 250 pc is 1\farcs5 at the distance of NGC 3256.)
The 1$\sigma$ errors of the NGC 253 data are smaller than the symbols.
\label{fig.coratio} }
\end{figure}

\acknowledgements
We appreciate the SMA staff for their skilled help before and during the observations.
We are also grateful to Barry Rothberg for kindly providing us with his near-IR image of the galaxy,
and to Kohji Tomisaka, Keiichi Wada, and Junichiro Makino for their helpful comments.
The presentation of the paper is improved by the comments from the referee.
The blue image in Fig. \ref{fig.opt} is based on photographic data obtained using The UK Schmidt Telescope.     
The UK Schmidt Telescope was operated by the Royal Observatory          
Edinburgh, with funding from the UK Science and Engineering Research    
Council, until 1988 June, and thereafter by the Anglo-Australian        
Observatory.  Original plate material is copyright (c) the Royal        
Observatory Edinburgh and the Anglo-Australian Observatory.  The        
plates were processed into the present compressed digital form with     
their permission.  The Digitized Sky Survey was produced at the Space   
Telescope Science Institute under US Government grant NAG W-2166.
This research has made use of the NASA/IPAC Extragalactic Database (NED),
and of NASA's Astrophysics Data System Bibliographic Services.
This research is based in part on observations made with the NASA/ESA Hubble Space Telescope, 
obtained from the data archive at the Space Telescope Institute. 
STScI is operated by the association of Universities for Research in Astronomy, Inc. under 
the NASA contract  NAS 5-26555.


\clearpage

\end{document}